\documentclass[reqno,12pt]{amsart}
\usepackage{fullpage}
\usepackage{amsfonts}
\usepackage{amssymb}
\usepackage{tikz}

\numberwithin{equation}{section}
\def\LRL/{Laplace-Runge-Lenz}

\def\Lagr{\mathcal{L}}
\def\Ham{\mathcal{H}}

\def\solnsp{{\mathcal E}}
\def\D{{\mathcal D}}
\def\Dt{\mathcal{D}_t}

\def\S{\mathcal{S}}

\def\intA{\mathcal{A}}
\def\intB{\mathcal{B}}

\def\X{{\bf X}}
\def\Y{{\bf Y}}
\def\pr{{\rm pr}}

\def\K{{\mathrm K}}
\def\F{{\mathrm F}}
\def\E{{\mathrm F}}

\def\const{{\rm const.}}
\def\Rnum{\mathbb{R}}

\def\Nnum{\mathbb{N}}

\def\sgn{{\rm sgn}}

\def\t{{\rm t}}
\def\det{\mathrm{det}}

\def\parder#1#2{\frac{\partial{#1}}{\partial{#2}}}

\def\parders#1#2#3{\frac{\partial^2{#1}}{\partial{#2}\partial{#3}}}

\newtheorem{prop}{Proposition}
\newtheorem{thm}{Theorem}

\def\Ref#1{Ref.\cite{#1}}

\begin{document}

\title{Noether symmetry groups, locally conserved integrals, and dynamical symmetries\\ in classical mechanics}

\author{
Stephen C. Anco
\\\lowercase{\scshape{
Department of Mathematics and Statistics, 
Brock University\\
St. Catharines, ON, Canada}} \\
}

\begin{abstract}
Several aspects of the connection between conserved integrals (invariants)
and symmetries are illustrated within a hybrid Lagrangian-Hamiltonian framework
for dynamical systems.
Three examples are considered: 
a nonlinear oscillator with time-dependent frequency
(one degree of freedom);
geodesics of a spheroid (two degrees of freedom);
Calogero-Moser-Sutherland system of interacting particles
(three degrees of freedom). 
For each system, a local generalization of Liouville integrability is shown.
Specifically, 
the variational point symmetries in a Lagrangian setting 
lead to corresponding locally conserved integrals 
which are found to commute in the Poisson bracket
imported from the equivalent Hamiltonian setting. 
Action-angle variables are then introduced in the Lagrangian setting,
which leads to explicit integration of the Euler-Lagrange equations of motion
locally in time.
\end{abstract}

\maketitle

\section{Introduction}\label{sec:intro}
\label{intro}

One of the pivotal results in classical mechanics is Noether's theorem
which provides a one-to-one correspondence between 
conserved integrals (invariants) and infinitesimal symmetries of
a Lagrangian or Hamiltonian variational principle.
Broadly, symmetries are transformations under which
the set of solution trajectories is mapped into itself. 
These transformations comprise two types:
\emph{point symmetries},
whose compositions hold independently of the equations of motion
and close in the space of configuration variables and the time variable; 
\emph{dynamical symmetries},
whose compositions close only on solution trajectories
and essentially involve the generalized velocities.

For a given dynamical system, 
all infinitesimal point symmetries can be determined by Lie's method,
whereas finding all infinitesimal dynamical symmetries requires 
knowledge of the general solution of the equations of motion 
\cite{Olv-book,BA-book}.
When a dynamical system possesses a Lagrangian or Hamiltonian variational principle,
infinitesimal symmetries that preserve the variational structure
constitute \emph{variational symmetries}.
In addition to their connection with conserved integrals, 
symmetries are also very important for understanding
physical and mathematical properties of conserved integrals and solution trajectories. 

A modern formulation of Noether's theorem for Lagrangian systems is known \cite{BA-book,Anc-review}
that uses only the variational structure of the equations of motion.
In particular, the correspondence between variational symmetries and conserved integrals
is stated without an explicit Lagrangian being needed.

Hamiltonian systems that possess
a sufficient number of globally conserved integrals commuting in the Poisson bracket
are known to be Liouville integrable in terms of action-angle variables
\cite{Arn-book,Gol.Poo.Saf-book}. 
Use of these variables allows the general solution of the equations of motion
to be obtained by integration.

In recent work \cite{Anc2026},
a hybrid Lagrangian-Hamiltonian framework has been developed 
which combines Noether's theorem and Liouville integrability
for any dynamical system with a variational principle. 
One highlight of this framework is that Liouville integrability is generalized to enable 
integration of the equations of motion locally in time
when the conservation of Poisson commuting conserved integrals 
holds only locally on solution trajectories.
In essence, the global conditions on conserved integrals
are relaxed so as to allow local integrability to hold. 

Locally conserved integrals differ from globally conserved integrals by having
conservation hold only piecewise on solution trajectories.
A well known example is the generalized \LRL/ vector in central force mechanics
\cite{Fra,Per,Anc.Mea.Pas},
which exhibits a jump whenever a particle reaches an up-coming periapsis point
on a solution trajectory that is precessing.
The time at which the particle reaches each periapsis is another example of
a conserved integral that is only conserved locally \cite{Anc.Mea.Pas},
even when a solution trajectory is non-precessing but closed,
as happens for the Kepler problem \cite{Anc.Gol}.

Any locally Liouville-integrable dynamical system automatically possess
a maximal set of locally conserved integrals.
Specifically, the number of functionally independent integrals
will be twice the number of configuration variables. 
Each locally conserved integral yields a corresponding variational symmetry
through Noether's theorem (in reverse).
These variational symmetries will generate a Lie group of symmetry transformations
on solutions of the equations of motion,
which comprise the Noether symmetry group of the system.

The purpose of the present paper is to explain and illustrate these results
in detail for three examples of dynamical systems:
\begin{enumerate}
\item
a nonlinear oscillator with time-dependent frequency
(with one degree of freedom);
\item 
geodesics of a spheroid
(with two degrees of freedom);
\item
Calogero-Moser-Sutherland system of interacting particles
(with three degrees of freedom). 
\end{enumerate}

A summary of the hybrid framework is presented in section~\ref{sec:framework},
with emphasis on local Liouville integrability.
The three examples are given in sections~\ref{sec:ex.0} to~\ref{sec:ex.2}.
Some concluding remarks are made in section~\ref{sec:remarks}.

\section{Summary of framework}\label{sec:framework}

To start, a brief review of variational principles will be given
for general dynamical systems
\begin{equation}\label{eom.gen}
  \ddot{q}^ i = f^i(t,q,\dot{q})
\end{equation}
with configuration variables (generalized positions) $q^i$, $i=1,\ldots,N$.

A system \eqref{eom.gen} admits an action principle
\begin{equation}\label{action.gen}
  S = \int_{t_1}^{t_2} \Lagr(t,q(t),\dot q(t))\, dt
\end{equation}
when the following relation holds, 
\begin{equation}\label{EL.eqn.gen}
  \frac{\delta\Lagr}{\delta q^i} = g_{ij} (f^j - \ddot{q}^j) =0
\end{equation}
where $\Lagr(t,q,\dot{q})$ is a Lagrangian,
and 
\begin{equation}\label{g.matrix}
  g_{ij} =  g_{ji} = \parders{\Lagr}{\dot{q}^i}{\dot{q}^j}
\end{equation}
is the Hessian matrix. 
It will be assumed that the Lagrangian is non-degenerate, namely $\det(g) \neq 0$.

An equivalent Hamiltonian formulation arises from the Legendre transformation 
\begin{equation}\label{Ham.gen}
  \Ham(t,p,q) = \big( p_i \dot{q}^i  - \Lagr(t,q,\dot{q}) \big)\big|_{\dot{q}=\dot{q}(q,p)}
\end{equation}
in terms of the canonical momenta variables
$p_i = \dfrac{\partial L}{\partial\dot{q}^i}$, $i=1,\dots,N$. 
Inverting this relation, 
which is ensured by the non-degeneracy assumption on the Hessian matrix \eqref{g.matrix},
allows changing variables from $(q^i,\dot{q}^i)$ to $(q^i,p_i)$,
yielding Hamilton's equations 
\begin{equation}\label{Ham.eom.gen}
  \dot{q}^i = \parder{\Ham}{p_i},
  \quad
  \dot{p}_i = -\parder{\Ham}{q^i}
\end{equation}
from the Lagrangian equations of motion $\dfrac{\delta\Lagr}{\delta q^i} = 0$. 

A Lagrangian is unique up to the addition of a total time-derivative,
$\Lagr \to \Lagr + \dot{{\mathcal A}}$ for any function $\mathcal A(t,q)$,
since this changes the action principle only by an irrelevant endpoint term.
In particular,
both $g_{ij}$ and $f^i$ are unchanged.
Correspondingly,
the Hamiltonian \eqref{Ham.gen} gets changed by a canonical transformation
$\tilde\Ham(t,\tilde p,q)  = \Ham(t,q,p)|_{p=p(\tilde p,q,t)} +\dfrac{\partial G(t,q,\tilde p)}{\partial t}$
arising from the generating function
$G = q^i \tilde p_i - \mathcal A$,
for any function $\mathcal A(t,q)$,
where $\tilde p_i = p_i + \dfrac{\partial\mathcal A}{\partial q^i}$, $i=1,\ldots,N$,
are the transformed canonical momenta. 
The resulting change of Hamiltonian variables $(q^i,p_i)$ to $(q^i,\tilde p_i)$
preserves the form of Hamilton's equations.

Let $\solnsp$ denote the solution space of trajectories $(q^i(t),\dot{q}^i(t))$
of the equations of motion.
This will be naturally viewed as a surface in the jet space, $J$,
with coordinates $(t,q^i,\dot{q}^i,\ddot{q}^i)$.
The main aspects of the hybrid Lagrangian-Hamiltonian framework
will be summarized next. 

\subsection{Variational symmetries and conserved integrals}

Infinitesimal variational symmetries represent generators of transformations
that leave the action principle invariant up to end point terms.
They are most naturally formulated in jet space by vector fields
\begin{equation}
  \X =P^i(t,q,\dot{q}) \partial_{q^i}, 
\end{equation}
which do not make any distinction between
point transformations and dynamical transformations,
as explained in \Ref{Anc2026}. 

The condition of symmetry invariance is given by 
\begin{equation}\label{inv.cond}
 \frac{\delta \pr\X(\Lagr)}{\delta q^i} =0
\end{equation}
since the variational derivative annihilates a function
if and only if the function is a total time derivative.
Symmetry invariance can be stated equivalently in terms of the equations of motion
\begin{equation}\label{inv.cond.eom}
  \frac{\delta \big(P^i g_{ij} (\ddot{q}^j -f^j)\big)}{\delta q^i} =0
\end{equation}
as shown in \Ref{Anc2026}.
This leads directly to a modern form of Noether's theorem. 

\begin{thm}\label{thm:noether}
For a dynamical system \eqref{EL.eqn.gen} possessing a variational principle, 
a function $C(t,q,\dot{q}^i)$ is a locally conserved integral $\dot{C}|_\solnsp =0$ 
if and only if 
the vector field $\X =P^i(t,q,\dot{q}) \partial_{q^i}$
is an infinitesimal variational symmetry \eqref{inv.cond}, 
where $C$ and $P^i$ are explicitly related via the Hessian matrix \eqref{g.matrix} by
\begin{equation}\label{PfromC.gen}
  P^ i = g^{-1}{}^{ij} C_{\dot{q}^j}
\end{equation}
and 
\begin{equation}\label{CfromP.gen}
  \begin{aligned}
  C(t,q,\dot{q})
  = & \int_{\mathcal{C}} \big(
  {- g_{ij}} f^i P^j + \dot{q}^k ( (g_{ij} f^i P^j)_{\dot{q}^k} + (g_{kj} P^j)_t + \dot{q}^i(g_{kj} P^j)_{q^i} ) \big)\,dt
  \\&\qquad
  -\big( (g_{ij} f^i P^j)_{\dot{q}^k} + (g_{kj} P^j)_t + \dot{q}^i(g_{kj} P^j)_{q^i} \big)\, dq^k
  + g_{kj} P^j \, d\dot{q}^k . 
  \end{aligned}
\end{equation}
Here $\mathcal C$ denotes any curve in the coordinate space $(t,q,\dot{q}^i)$, 
starting at an arbitrary point $(t_0,q^i_0,\dot{q}^i_0)$. 
\end{thm}

For any locally conserved integral $C(t,q,\dot{q})$,
the corresponding infinitesimal variational symmetry
preserves the extremals of the action principle
and thereby maps the solution space $\solnsp$ into itself.
The resulting symmetry on solutions is given by the prolonged vector field 
\begin{equation}\label{varsym.solnsp.gen}
  \X^\solnsp_{(C)} =
  g^{-1}{}^{ij}\partial_{\dot{q}^j} C \partial_{q^i}
  + \Dt(g^{-1}{}^{ij}\partial_{\dot{q}^j})\partial_{\dot{q}^i}
\end{equation}
where
\begin{equation}\label{time.der.solnsp}
\Dt
= \partial_t + \dot{q}^i\partial_{q^i} + f^i\partial_{\dot{q}^i}
\end{equation}
represents the time derivative restricted to $\solnsp$.

\subsection{Poisson bracket and its uses} 

The set of all locally conserved integrals of a dynamical system \eqref{EL.eqn.gen}
is a linear space,
since any linear combination of locally conserved integrals
is again locally conserved.
An arbitrary (smooth) function of any locally conserved integral is also 
locally conserved.

This space has two main properties, which can be stated
in terms of the Poisson bracket.
For this purpose,
it will be useful to transcribe the Poisson bracket
from Hamiltonian variables to Lagrangian variables: 
\begin{equation}\label{Lagr.PB}
  \{F_1,F_2\}  = \nabla^\t F_1\, \mathbf{J}\, \nabla F_2
  =   g^{-1}{}^{ij}  \Big( \parder{F_1}{q^i}\parder{F_2}{\dot{q}^j} -\parder{F_2}{q^i}\parder{F_1}{\dot{q}^j} \Big)
  +  c^{ij}  \parder{F_1}{\dot{q}^i}\parder{F_2}{\dot{q}^j} 
\end{equation}
for any two functions $F_1$ and $F_2$ of $t$, $q$, $\dot{q}$, 
where 
\begin{equation}\label{h.matrix}
  c^{ij} = - c^{ji} = g^{-1}{}^{ik}g^{-1}{}^{jl} (h_{ij} - h_{ji}),
  \quad
  h_{ij} = \parders{\Lagr}{q^i}{\dot{q}^j} . 
\end{equation}
Here $\nabla = \begin{pmatrix} \partial_{q^i} \\ \partial_{\dot{q}^i} \end{pmatrix}$
denotes the gradient in Lagrangian coordinates;
$\t$ denotes the transpose;
and 
\begin{equation}\label{PB.symplectic}
\mathbf{J} = 
\begin{pmatrix}
  0 & g^{-1}{}^{ij} \\ - g^{-1}{}^{ij} & c^{ij}
\end{pmatrix}
\end{equation}
denotes the symplectic matrix which is skew and obeys the Jacobi identity.

Recall that a function $C(t,q,\dot{q})$ is a locally conserved integral
if and only if
\begin{equation}\label{local.conserved.Ham}
  \parder{C}{t} + \{ C, \Ham \} =0 . 
\end{equation}
A \emph{constant of motion} is a conserved integral
satisfying $\dfrac{\partial C}{\partial t}=0$,
and otherwise when a conserved integral contains $t$ explicitly,
$\dfrac{\partial C}{\partial t}\neq 0$,
it is a \emph{temporal integral of motion}.
Two locally conserved integrals $C_1$ and $C_2$
are \emph{functionally independent}
if $k(C_1,C_2)=0$ holds only for the trivial function $k\equiv0$.
The counterpart condition for variational symmetries is that 
$\X^\solnsp_{(C_1)}$ and $\X^\solnsp_{(C_2)}$ 
are \emph{independent over the solution space}
if and only if 
\begin{equation}\label{X.dependent}
  k_{C_1} \X^\solnsp_{(C_1)} + k_{C_2} \X^\solnsp_{(C_2)} =0
\end{equation}
holds only when the function $k(C_1,C_2)\equiv0$ is trivial. 
This is a stronger condition than linear independence,
since two variational symmetries could be linearly independent, 
$c_1\X^\solnsp_{(C_1)} + c_2\X^\solnsp_{(C_2)} \neq 0$ for all constants $c_1$ and $c_2$,
but still be dependent over the solution space \eqref{X.dependent}.

Now, the first property of the linear space of conserved integrals
is that it carries an action of every variational symmetry. 

\begin{thm}\label{thm:X.PB.C1.C2}
Let $C_1(t,q,\dot{q})$ and $C_2(t,q,\dot{q})$ 
be locally conserved integrals. 
Their corresponding infinitesimal variational symmetries \eqref{varsym.solnsp.gen}
projected into the solution space $\solnsp$
satisfy the relation 
\begin{equation}\label{X.PB.C1.C2}
 \X^\solnsp_{(C_1)}(C_2)
  = -\X^\solnsp_{(C_2)}(C_1)
  = \{ C_2, C_1 \} . 
\end{equation}
\end{thm}

The second property is that the linear space of conserved integrals
has the structure of a Lie algebra under the Poisson bracket. 

\begin{thm}\label{thm:C1.C2.varsymm.PB} 
For any conserved integrals $C_1(t,q,\dot{q})$, $C_2(t,q,\dot{q})$,
the commutator of their corresponding infinitesimal variational symmetries \eqref{varsym.solnsp.gen}
projected into the solution space $\solnsp$ is given by 
 \begin{equation}\label{C1.C2.varsymm.PB}
    [\X^\solnsp_{(C_2)},\X^\solnsp_{(C_1)}] = \X^\solnsp_{(\{ C_1, C_2 \})} . 
 \end{equation}
\end{thm}

A main corollary of this theorem is that
there is a homomorphism of the Poisson bracket Lie algebra
onto the Lie algebra of all variational symmetries
where the Lie bracket is given by the commutator of vector fields. 
The kernel of this homomorphism consists of constants, which are trivially conserved.
They are mapped into the zero symmetry by the Noether correspondence \eqref{PfromC.gen}.

Theorems~\ref{thm:X.PB.C1.C2} and~\ref{thm:C1.C2.varsymm.PB}
have an important computational use. 
If the commutators of a set of infinitesimal symmetries are known,
then relation \eqref{C1.C2.varsymm.PB} directly yields
the corresponding Poisson brackets. 
Alternatively, if the Poisson brackets are known,
then relation \eqref{C1.C2.varsymm.PB} provides
the commutators of the corresponding infinitesimal symmetries. 
In addition, 
the Poisson brackets also determine 
the action of the corresponding infinitesimal symmetries on the conserved integrals
through relation \eqref{X.PB.C1.C2}. 
Instead if the symmetry actions are known, 
then they directly yield the Poisson brackets. 
This is particularly useful when the symmetries have a geometrical meaning
which allow their action to be determined entirely by geometric considerations.

The following result summaries the general properties of variational symmetries. 

\begin{prop}\label{prop:count}
For a dynamical system \eqref{EL.eqn.gen} with $N$ degrees of freedom:
(i) The number of functionally-independent locally conserved integrals
is $2N$,
which is equal to the number of variational symmetries that are independent over the solution space.
(ii) The total number of linearly independent variational symmetries is infinite.
(iii) A set of locally conserved integrals is homomorphic to a finite-dimensional Lie algebra of variational symmetries
if and only if their Poisson brackets close linearly.
\end{prop}

\subsection{Symmetry transformation groups}

Every infinitesimal symmetry 
whether it is variational or not,
generates a one-parameter Lie group of symmetry transformations 
that maps the solution space of the equations of motion into itself. 
This mapping is given by the integral curve of the vector field \eqref{varsym.solnsp.gen}
via the exponential mapping 
\begin{equation}\label{X.transformation.group}
  (q^i,\dot{q}^i)_\solnsp   \to  (q^i{}^*,\dot{q}^i{}^*)_\solnsp
  = \exp(\varepsilon\,\X^\solnsp) (q^i,\dot{q}^i)_\solnsp
\end{equation}
with parameter $\varepsilon$,
where $\varepsilon=0$ yields the identity transformation.
The infinitesimal form of the transformation group is simply 
\begin{equation}
  (q^i{}^*,\dot{q}^i{}^*)_\solnsp
  = (q^i,\dot{q}^i)_\solnsp +\varepsilon \big(P^i(t,q,\dot{q})_\solnsp ,\Dt P^i(t,q,\dot{q})_\solnsp \big) + O(\varepsilon^2) . 
\end{equation}

An equivalent transformation group can be defined
in the coordinate space $(t,q^i,\dot{q}^i)$
by exponentiating the extended vector field 
\begin{equation}\label{Y.gen}
  \Y = \X^\solnsp + \tau \D_t 
  = \tau\partial_{t} +\big(P^i + \tau\dot{q}^i\big)\partial_{q^i} + \big(\D_tP^i +\tau f^i\big)\partial_{\dot{q}^i}
\end{equation}
where $\tau$ is a completely arbitrary function of $t$, $q^i$, $\dot{q}^i$.
In particular, 
the vector fields $\X^\solnsp$ and $\Y$ have the same action
on the solution space $\solnsp$, 
as shown by 
\begin{equation}\label{X.C.Y.C}
  \X^\solnsp(C)|_\solnsp = \Y(C)|_\solnsp , 
\end{equation}
which holds for any locally conserved integral $C(t,q,\dot{q})$. 
The integral curve of the vector field \eqref{Y.gen}
projected into the solution space $\solnsp$ 
thereby yields the Lie group of transformations
\begin{equation}\label{Y.transformation.group} 
  (t,q^i,\dot{q}^i)_\solnsp \to  (t^\dagger,q^i{}^\dagger,\dot{q}^i{}^\dagger)_\solnsp
  = \exp(\varepsilon\,\Y) (t,q^i,\dot{q}^i)_\solnsp
\end{equation}
with parameter $\varepsilon$.
For any choice of $\tau(t,q,\dot{q})$,
this provides an equivalent representation of the symmetry transformation group \eqref{X.transformation.group} generated by $\X^\solnsp$. 
Consequently, $\tau$ constitutes a gauge freedom in the description of
symmetry transformation groups. 
This will turn out to enable finding an explicit form of
the symmetry transformation group for any infinitesimal variational symmetry. 

There are two different types of variational symmetry groups: 
\emph{point symmetries} and \emph{dynamical symmetries}.
They are distinguished by how $P^i(t,q,\dot{q})$ depends on $\dot{q}^i$.

A point symmetry arises when (and only when)
\begin{equation}\label{point.symm}
  P^i = \eta^i(t,q) -\tau(t,q) \dot{q}^i
\end{equation}
is linearly proportional to $\dot{q}^i$,
where $\tau(t,q)$ is identified with the gauge function
in the extended vector field \eqref{Y.gen}. 
This yields a transformation group \eqref{Y.transformation.group} with the form 
\begin{equation} 
(t,q^i,\dot{q}^i)\to (t^\dagger,q^i{}^\dagger,\dot{q}^i{}^\dagger)  =
  (t,q^i,\dot{q}^i)
  + \varepsilon \big( \tau(t,q),\eta^i(t,q), \dot{\eta}^i(t,q,\dot{q}) -\dot{\tau}(t,q,\dot{q})\dot{q}^i \big)
  + O(\varepsilon^2) , 
\end{equation}
which is the prolongation to $(t,q^i,\dot{q}^i)$ of a point transformation on $(t,q^i)$.
In particular, the infinitesimal point transformation is given by the vector field
\begin{equation}\label{Y.point.symm}
  \Y= \tau(t,q)\partial_t + \eta^i(t,q)\partial_{q^i} . 
\end{equation}

When a vector field has a more general form,
\begin{equation}\label{dyn.symm}
  \parder{P^i}{\dot{q}^j} \neq -\tau(t,q)  \delta_j{}^i,
\end{equation}
then it instead describes a dynamical symmetry. 
In general there is no gauge choice that will allow
the resulting transformation group \eqref{Y.transformation.group}
to be expressed as a point transformation on $(t,q^i,\dot{q}^i)$.
Equivalently, the transformation of configuration variables $q^i$
will explicitly involve the velocity variables $\dot{q}^i$.

\subsection{Locally Liouville integrable systems}

A dynamical system \eqref{EL.eqn.gen} is locally Liouville integrable
if it possesses $N$ locally conserved integrals $C_i(t,q,\dot{q})$, $i=1,\dots, N$,
that are functionally independent and whose Poisson brackets vanish, $\{C_i,C_j\}=0$.
Action-angle variables can then be introduced
which yield $N$ additional conserved integrals. 
As explained in \Ref{Anc2026},
a purely Lagrangian formulation of this local integrability can be stated. 

\begin{thm}\label{thm:liouville}
For a dynamical system \eqref{EL.eqn.gen} that is locally Liouville integrable, 
let 
\begin{equation}\label{S.action.angle}
\S(t,q,C) = \int \parder{\Lagr}{\dot{q}^j}(t,q,\dot{q}(t,q,C))\, dq^j
\end{equation}
where $\dot{q}^j(t,q,C)$ is given by inverting $C_i=C_i(t,q,\dot{q})$. 
The generalized angle variables 
\begin{equation}\label{action.angle.variables}
\Theta^i = \parder{\S}{C_i} = \int g_{jk} \partial_{C_i}\dot{q}^k(t,q,C)\, dq^j
\end{equation}
lead to $N$ additional conserved integrals,
where $g_{jk}$ is the Hessian matrix \eqref{g.matrix}.
In particular:\\
(i) if the Lagrangian is autonomous, $\partial_t \Lagr =0$,
then by choosing
\begin{equation}
  C_1 = H = \dot{q}^i \partial_{\dot{q}^i}\Lagr -\Lagr,
\end{equation}
the generalized angles $\Theta^2,\ldots, \Theta^N$ constitute $N-1$ constants of motion,
and $T= t - \Theta^1$ is a temporal integral of motion.\\
(ii) If the Lagrangian is non-autonomous, $\partial_t \Lagr \neq0$,
then 
\begin{equation}\label{liouville.T}
\Upsilon^i=  \int (\partial_t\Theta^i + \partial_{q^j}\Theta^i\dot{q}^j(t,q,C))\,dt -\Theta^i
\end{equation}
are integrals of motion, 
in terms of which $N-1$ constants of motion are given by
solutions of the linear homogeneous partial differential equation 
$\dfrac{\partial K}{\partial C_i} \partial_{\Upsilon^i} F(\Upsilon^1,\ldots,\Upsilon^N)=0$ 
where $K = \dot{\S}(t,q,C)  -\Lagr(t,q,\dot{q}(t,q,C))$. 
\end{thm}

The $N$ locally conserved integrals $C_i$ yield $N$ variational symmetries
\begin{subequations}\label{liouville.C.X}
\begin{equation}
\X^\solnsp_{(C_i)} = P^j_{(C_i)} \partial_{q_j} + \Dt P^j_{(C_i)} \partial_{\dot{q}_j}
\end{equation}
where
\begin{equation} 
P^j_{(C_i)} = g^{-1}{}^{jk}\partial_{\dot{q}^k} C_i .  
\end{equation}
\end{subequations}
These variational symmetries are independent over the solution space 
and mutually commute since $\{C_j,C_i\} =0$.

The additional $N$ locally conserved integrals given by Theorem~\ref{thm:liouville}
similarly yield $N$ variational symmetries.
Their mutual commutators as well as their commutators with the variational symmetries \eqref{liouville.C.X}
are given by variational symmetries that arise from the corresponding Poisson brackets.

\section{Nonlinear oscillator with time-dependent frequency}\label{sec:ex.0}

The governing equation of a nonlinear oscillator with a time-dependent frequency $\omega(t)$
is given by 
\begin{equation}\label{ex0.q.eqn}
  \ddot{q} + \omega(t)^2 q + f(q) =0
\end{equation}
where $f'\not\equiv 0$.
This equation arises from the Lagrangian 
\begin{equation}
\Lagr = \tfrac{1}{2} (\dot{q}^2 - \omega(t)^2 q^2) -\int f(q)\, dq
\end{equation}
and has Hamilton variables $(q,p)$ where $p=\dot{q}$. 

As a preliminary aim,
all nonlinearities $f(q)$ will be found
for which there exists a conserved integral generalizing
the well-known Lewis invariant \cite{Lew} in the case of a linear oscillator.

\subsection{Point symmetries and conserved integrals}

The determining condition \eqref{inv.cond}
for variational point symmetries \eqref{point.symm}
gives the system of equations
\begin{subequations}
\begin{gather}
  \xi_{qq} = 0,
  \quad
  q\xi_t + 2\eta =0,
  \quad
  q\eta_{q} -\eta =0,
\label{ex0.deteqn1}  
  \\
  \eta_{tt}  +2q \omega(t) \omega'(t) \xi + (4\omega(t)^2 + 3f(q)/q + f'(q)) \eta =0 , 
\label{ex0.deteqn2}  
  \\
  f''(q) +3 f'(q)/q - 3 f(q)/q^2 =0 ,
\label{ex0.f.deteqn}
\end{gather}
\end{subequations}
for $\xi(t,q)$, $\eta(t,q)$, and $f(q)$.
Solving equation \eqref{ex0.f.deteqn},
with the condition $f''\not\equiv0$,
determines $f(q) = \alpha q + \beta q^{-3}$,
where $\alpha$ and $\beta$ are arbitrary constants.
The first term can be absorbed into the quadratic term in the Lagrangian
via $\omega^2 \to \omega^2 -\alpha$. 
A scaling transformation $q\to \lambda q$
allows setting $|\beta|$ to be any desired (non-zero) value,
whereby $\beta^2 = 1$ without loss of generality. 
Then the general solution of the remaining equations \eqref{ex0.deteqn1}--\eqref{ex0.deteqn2}
is given by
\begin{equation}\label{ex0.xi.eta}
  \xi = \sigma(t),
  \quad
  \eta = \tfrac{1}{2} \sigma'(t) q , 
\end{equation}
where $\sigma(t)$ satisfies  the differential equation
\begin{equation}
  \sigma''' + 4\omega(t)^2 \sigma' + 4\omega(t)\omega'(t) \sigma =0 . 
\end{equation}
Use of the integrating factor $\sigma^2$ yields leads to the reduced differential equation
\begin{equation}\label{ex0.sigma.c.eqn}
  \sigma'' + \omega(t)^2 \sigma - c\sigma^{-3} =0,
  \quad
  c=\const . 
\end{equation}

Hence, the nonlinearity for which a variational point symmetry exists is 
\begin{equation}\label{ex0.f}
  f(q) = \beta q^{-3},
  \quad
  \beta^2=1 . 
\end{equation}
This gives the nonlinear oscillator equation 
\begin{equation}\label{ex0.eom}
  \ddot{q} + \omega(t)^2 q +\beta q^{-3} =0 , 
\end{equation}
with the Lagrangian
\begin{equation}\label{ex0.Lagr}
\Lagr = \tfrac{1}{2} (\dot{q}^2 - \omega(t)^2 q^2 +\beta q^{-2}) . 
\end{equation}
Its variational point symmetry, given by the components \eqref{ex0.xi.eta},
is the vector field 
\begin{equation}\label{ex0.pointsymm}
  \X =   \sigma(t) \dot{q}\partial_t + \tfrac{1}{2} \sigma'(t) q\partial_q 
\end{equation}
in the coordinate space $(t,q)$.

This variational point symmetry has an equivalent formulation \eqref{point.symm}
given by 
\begin{equation}\label{ex0.varsymm}
  \X =  \big( \tfrac{1}{2} \sigma'(t) q - \sigma(t) \dot{q} \big)\partial_q , 
\end{equation} 
which is useful for applying Noether's theorem (cf Theorem~\ref{thm:noether}). 
Through the Noether correspondence \eqref{PfromC.gen},
this symmetry yields the conserved integral
\begin{equation}
  C = \tfrac{1}{2}\big(
  (\sigma(t) \dot{q}- \sigma'(t) q)^2   + c \sigma(t)^{-2} q^2 -\beta \sigma(t)^2 q^{-2} \big)
\end{equation}
where $c$ is an arbitrary constant
and $\sigma(t)$ is any solution of the differential equation \eqref{ex0.sigma.c.eqn}.
Since $c$ can be chosen freely,
$C$ constitutes a family of conserved integrals.
Each one clearly is single-valued and hence is globally conserved. 
The form of $C$ is a generalization of the well-known Lewis invariant \cite{Lew}
\begin{equation}
  I = \tfrac{1}{2}\big(  (\rho(t) \dot{q}- \rho'(t) q)^2   +  \rho(t)^{-2} q^2 \big)
\end{equation}
which exists for a linear oscillator $\ddot{q} +\omega(t)^2 q=0$,
where $\rho(t)$ satisfies the Ermakov-Pinney equation 
$\ddot{\rho} + \omega(t)^2 \rho =\rho^{-3}$.
A symmetry derivation of this invariant can be found in \Ref{Rog.Ram,Shi.Mui.Lea}. 

Notice that for $c=0$ the differential equation \eqref{ex0.sigma.c.eqn} becomes
\begin{equation}\label{ex0.sigma.eqn}
  \sigma'' + \omega(t)^2 \sigma =0 , 
\end{equation}
which is a linear oscillator with time-dependent frequency.
Each solution yields a conserved integral
\begin{equation}\label{ex0.C}
  C = \tfrac{1}{2}\big(
  (\sigma(t) \dot{q}- \sigma'(t) q)^2   -\beta \sigma(t)^2 q^{-2} \big)
\end{equation}
for the nonlinear oscillator equation \eqref{ex0.eom}.
This will be used for the subsequent analysis,
with $\sigma(t)$ being any fixed (non-zero) solution of
the linear differential equation \eqref{ex0.sigma.eqn}.

\subsection{Liouville integrability}

The conserved integral \eqref{ex0.C} implies that
the nonlinear oscillator equation \eqref{ex0.eom} is locally Liouville integrable. 
An action-angle variable \eqref{liouville.T}
comes from the integral expression \eqref{S.action.angle} given by 
\begin{equation}
  \S = \int \parder{\Lagr}{\dot{q}}(t,q,C)\, dq
  = \int \dot{q}(t,q,C)\, dq
\end{equation}
where, from expression \eqref{ex0.C}, 
\begin{equation}\label{ex0.dotq}
  \dot{q} =  \sigma(t)^{-1}\big(
  \sigma'(t) q   \pm \sqrt{2C +\beta  \sigma(t)^2 q^{-2}}
  \big) . 
\end{equation}
Taking the derivative of $\S$ with respect to $C$ yields
\begin{equation}\label{ex0.angle.C}
  \Theta(t,q,C):= \parder{\S}{C}
  = \int \frac{\pm  \sigma(t) q^2}{\sqrt{2Cq^2 +\beta \sigma(t)^2}}\,dq
\end{equation}
where 
\begin{equation}
\parder{\Theta}{t} + \parder{\Theta}{q} \dot{q}(t,q,C)
= \sigma(t)^{-2} . 
\end{equation}
The action-angle variable is then given by 
\begin{equation}\label{ex0.Upsilon.t}
  \Upsilon(t,q,C): = \int \Big(\parder{\Theta}{t} + \parder{\Theta}{q} \dot{q}(t,q,C)\Big)\, dt - \Theta
  = \int \sigma(t)^{-2}\,dt -\int \frac{\pm  \sigma(t) q^2}{\sqrt{2Cq^2 +\beta \sigma(t)^2}}\,dq . 
\end{equation}
It will represent a locally conserved integral 
provided that the dynamical features of the oscillator motion
are used to choose the integration constant in both of the integral terms. 
There are two natural choices for the integration constant in the $q$-integral term: 
a turning point  where $\dot{q}=0$,
or an inertial point where $\ddot{q}=0$.
Once a choice is fixed, then the conserved integral \eqref{ex0.Upsilon.t}
has the meaning 
\begin{equation}\label{ex0.Upsilon.T}
  \Upsilon(T,q,C) = \int \sigma(t)^{-2}\,dt\big|_{t=T}
\end{equation}
where $T$ is the time at which $q(t)$ reaches the chosen point.
Although $\Upsilon$ will still depend on an integration constant in the $t$-integral term,
this constant will cancel out when
expressions \eqref{ex0.Upsilon.T} and \eqref{ex0.Upsilon.t} are equated. 

From the oscillator equation \eqref{ex0.eom},
inertial points  $q=q^*$ are determined by $q^*{}^4 =-\beta \omega(t)^{-2}$,
which requires $\beta =-1$.
This gives
\begin{equation}\label{ex0.ip}
  |q^*| = \sqrt{\omega(t)}^{-1} ,
  \quad
  \beta =-1 . 
\end{equation}
Turning points $q=q_*$ are the roots of the right-hand side of expression \eqref{ex0.dotq}.
Squaring it yields a quadratic equation in $q_*^2$, which gives
\begin{align}
  &  |q_*| = \frac{\sqrt{C +\sqrt{C^2 +\beta \sigma'(t)^2 \sigma(t)^2}}}{|\sigma'(t)|},
  \quad
  \beta = \pm 1 , 
  \label{ex0.tp1}
  \\
  & |q_*| = 
  \quad
    \frac{\sqrt{C -\sqrt{C^2 +\beta \sigma'(t)^2 \sigma(t)^2}}}{|\sigma'(t)|},\
    \beta = -1 , 
  \label{ex0.tp2}
\end{align}
where the sign in expression \eqref{ex0.dotq} must be chosen as
$-\sgn(\sigma(t) \sigma'(t))$.

Thus, the following conserved integrals \eqref{ex0.Upsilon} are obtained.
Let
\begin{equation}
A = \sqrt{1+ \beta C^{-1} \sigma(t)^2 \sigma'(t)^2 }
\end{equation}
and
\begin{equation}
  \intB(t) = \int \sigma(t)^{-2}\,dt . 
\end{equation}
First, for $\beta=1$, 
there is a pair of turning points \eqref{ex0.tp1},
giving
\begin{subequations}\label{ex0.Upsilon}
\begin{equation}\label{ex0.betais1.Upsilon.tp}
  \Upsilon = \intB(T)
  =   \intB(t) + \tfrac{1}{2}\sgn(\sigma(t) \sigma'(t)) \big(
  C^{-1} \sqrt{2C\sigma(t)^{-2} q^2 +1} 
  -\tfrac{1}{2}(1 + A) |\sigma(t)\sigma'(t)|^{-1} \big) . 
\end{equation}
Second, for $\beta=-1$,
there are two pairs of turning points \eqref{ex0.tp1} and \eqref{ex0.tp2},
giving
\begin{equation}\label{ex0.betais-1.Upsilon.tp}
  \Upsilon = \intB(T)
  = \intB(t)   + \tfrac{1}{2}\sgn(\sigma(t) \sigma'(t)) \big(
  C^{-1} \sqrt{2C\sigma(t)^{-2} q^2 -1}
  -\tfrac{1}{2}(1 \pm A) |\sigma(t)\sigma'(t)|^{-1} \big) . 
\end{equation}
In addition, there is a pair of inertial points \eqref{ex0.ip},
giving
\begin{equation}\label{ex0.betais-1.Upsilon.ip}
  \Upsilon = \intB(T)
  = \intB(t)   \pm \tfrac{1}{2} C^{-1} \big(
    \sqrt{2C\sigma(t)^{-2} \omega(t)^{-1} -1} - \sqrt{2C\sigma(t)^{-2} q^2 -1} \big) , 
\end{equation}
\end{subequations}
where the $\pm$ sign here is same as in expression \eqref{ex0.dotq}.

The behaviour of $\Upsilon$ and $T$ will be discussed next.
Firstly, the right-hand side of $\Upsilon$ is manifestly single-valued,
and hence the quantity $\intB(T)$ is globally conserved.
Secondly, $\intB(T)$ is oscillatory in $T$, due to $\sigma(t)$ being oscillatory in $t$,
which implies the quantity $T$ is multi-valued and therefore it is only locally conserved.
Importantly, each of the expressions \eqref{ex0.Upsilon} for $\Upsilon$ 
can be solved algebraically for $q$ as a function of $t$,
which thus provides the explicit solution of the nonlinear oscillator equation \eqref{ex0.eom},
parameterized in terms of $T$ and $C$ (instead of initial values).

\subsection{Dynamical symmetries and Poisson brackets}

Each of the locally conserved integrals \eqref{ex0.betais1.Upsilon.tp}--\eqref{ex0.betais-1.Upsilon.ip}
yields an infinitesimal symmetry of the Lagrangian \eqref{ex0.Lagr}
through the correspondence \eqref{PfromC.gen} in Noether's' theorem:
\begin{equation}\label{ex0.X.Upsilon}
  \X_{(\Upsilon)} = P_{(\Upsilon)}\partial_q . 
\end{equation}
Since these conserved integrals contain $C$, which depends nonlinearly on $\dot{q}$, 
the symmetries will be dynamical.
Hereafter,
their dependence on $\dot{q}$ will be substituted in terms of $C$
through equation \eqref{ex0.dotq}.
For $\beta=1$: 
\begin{equation}\label{ex0.P.tp1}
  P_{(\Upsilon)} = 
  \tfrac{1}{2}C^{-1} \big( 
  q + C^{-1} q^{-1}\sigma(t)^2\big( 1 - \sqrt{2C \sigma(t)^{-2} q^2 + 1}\,\sqrt{C^2 (\sigma(t) \sigma'(t))^{-2} +1}^{-1}\big)
  \big)  . 
\end{equation}
For $\beta=-1$:
\begin{equation}\label{ex0.P.tp2}
  P_{(\Upsilon)} =
  \tfrac{1}{2} C^{-1} \big( 
  q - C^{-1} q^{-1} \sigma(t)^2\big(1   \mp \sqrt{2C \sigma(t)^{-2} q^2 - 1}\,\sqrt{C^2 (\sigma(t) \sigma'(t))^{-2} -1}^{-1} \big) 
  \big) , 
\end{equation}
and
\begin{equation}\label{ex0.P.ip}
  P_{(\Upsilon)}
  = \tfrac{1}{2} C^{-1} \big( 
  q -C^{-1} q^{-1} \big( \sigma(t)^2   + (C\omega(t)^{-1}-\sigma(t)^2) \sqrt{2 C q^2 -\sigma(t)^2} \,\sqrt{2C\omega(t)^{-1} -\sigma(t)^2}^{-1} \big)
  \big) . 
\end{equation}

The infinitesimal point symmetry \eqref{ex0.varsymm} can be expressed similarly:
\begin{equation}\label{ex0.X.C}
  \X_{(C)} =P_{(C)}\partial_q,
  \quad
P_{(C)}(t,q,C) =  \mp q^{-1} \sqrt{2C q^2 + \beta \sigma(t)^2} |\sigma(t)| . 
\end{equation}
This form is useful for obtaining the action of this symmetry on the conserved integrals,
which will yield their Poisson brackets with $C$ through Theorem~\ref{thm:X.PB.C1.C2}.

By direct computation, 
$\X^\solnsp_{(C)} \Upsilon = P_{(C)}\partial_q\Upsilon + (\X^\solnsp_{(C)} C) \partial_{C}\Upsilon$
using chain rule. 
Next observe that $\X^\solnsp_{(C)} C=0$ due to relation \eqref{X.PB.C1.C2},
whereby
$\X^\solnsp_{(C)}\Upsilon = P_{(C)}\partial_q\Upsilon$.
After simplifications,
this yields the symmetry action 
\begin{equation}
\X^\solnsp_{(C)}\Upsilon =1 , 
\end{equation}
and hence 
\begin{equation}\label{ex0.X.Upsilon.C}
\X^\solnsp_{(\Upsilon)}C =-1
\end{equation}
from relation \eqref{X.PB.C1.C2}. 

Now, the second equality shown in relation \eqref{X.PB.C1.C2}
directly gives the Poisson bracket 
\begin{equation}
\{\Upsilon,C\} =1 . 
\end{equation}
This implies, from Theorem~\ref{thm:C1.C2.varsymm.PB}, that 
\begin{equation}\label{ex0.commutator}
  [\X^\solnsp_{(\Upsilon)},\X^\solnsp_{(C)}] =0 . 
\end{equation}
Thus, the point symmetry commutes with dynamical symmetry.

\subsection{Noether symmetry group}

The Noether symmetry group of the nonlinear oscillator \eqref{ex0.eom}
is a two-dimensional Lie group of transformations
generated by the infinitesimal point symmetry 
and the infinitesimal dynamical symmetry \eqref{ex0.X.Upsilon}.
This group is abelian, due to the vanishing commutator \eqref{ex0.commutator}. 

The point symmetry transformations have two equivalent forms.
First, the vector field \eqref{ex0.pointsymm} defining the infinitesimal symmetry
represents a flow
$(t,q) \to (t^\dagger,q^\dagger)$
given by
\begin{equation}
  \frac{d t^\dagger}{d\varepsilon} = \sigma(t^\dagger),
  \quad
  \frac{d q^\dagger}{d\varepsilon} = \tfrac{1}{2} \sigma'(t^\dagger) q^\dagger ,
\end{equation}
with parameter $\varepsilon\in\Rnum$,
where $\varepsilon=0$ gives the identity transformation.
These differential equations are easy to integrate explicitly,
yielding
\begin{equation}\label{ex0.point.t.q.group}
  \int_t^{t^\dagger} \sigma(y)^{-1}\, dy = \varepsilon,
  \quad
  q^\dagger = \sqrt{|\sigma(t^\dagger)|/|\sigma(t)}\, q . 
\end{equation}
The transformation on $t$ is only implicit.

Alternatively, an explicit form of the point transformation arises
from the exponential mapping \eqref{X.transformation.group}
which uses the infinitesimal transformation in the form \eqref{ex0.X.C}
prolonged to the space $(q,\dot{q})$.
This prolongation is most simply given by working in the space $(q,C)$,
where $\dot{q}$ is substituted in terms of $C$ through equation \eqref{ex0.dotq}.
Since $\X^\solnsp_{(C)}C=0$ as noted earlier, $C$ is an invariant, 
from which the transformation $\dot{q} \to \dot{q}^\dagger$
can be obtained straightforwardly in terms of the transformation $q \to q^\dagger$. 
The latter is given by solving the differential equation 
\begin{equation}
  \frac{d q^\dagger}{d\varepsilon} 
  = \mp q^\dagger{}^{-1} \sqrt{2C q^\dagger{}^2 + \beta \sigma(t)^2} |\sigma(t)| 
\end{equation}
with $q^\dagger|_{\varepsilon=0} = q$.
This yields
\begin{equation}\label{ex0.point.q.group}
  q^\dagger =q \sqrt{ 
  1 \mp 2 |\sigma(t)| q^{-2} \sqrt{2C q^2 +\beta \sigma(t)^2} \varepsilon 
  + 2 C |\sigma(t)|^2 q^{-1} \varepsilon^2 }
\end{equation}
where the $\mp$ sign is the negative of the sign in the expression \eqref{ex0.dotq}
for $\dot{q}$.
Then equation \eqref{ex0.dotq} for $\dot{q}$
combined with expression \eqref{ex0.C} for $C$
gives 
\begin{equation}\label{ex0.point.dotq.group}
  \dot{q}^\dagger =
  (\dot{q} -\sigma'(t)\sigma(t)^{-1}q)
  \sqrt{1 + \beta\sigma(t)^2 (q^\dagger{}^{-2} -q^{-2})(\sigma(t)^2 \dot{q} -\sigma'(t) q)^{-1}}
  + \sigma'(t) \sigma(t)^{-1} q . 
\end{equation}
This explicit transformation \eqref{ex0.point.q.group}--\eqref{ex0.point.dotq.group} on $(q,\dot{q})$ 
is equivalent to the implicit transformation \eqref{ex0.point.t.q.group} on $(t,q)$.

The dynamical symmetry transformations arise from the flow
defined by the vector field
\begin{equation}\label{ex0.X.solnsp}
  \X_{(\Upsilon)}^\solnsp = P_{(\Upsilon)}(t,q,C) \partial_q + \Dt P_{(\Upsilon)}(t,q,C) \partial_{\dot{q}} 
\end{equation}    
in the space $(q,\dot{q})$, 
where $\Dt$ is the time derivative restricted to the space of solutions of the equations of motion \eqref{ex0.eom}.   
However, it will be simpler to use the flow in the space $(q,C)$,
similarly to the previous method for the point symmetry \eqref{ex0.point.q.group}--\eqref{ex0.point.dotq.group}.
This will involve utilizing the symmetry action \eqref{ex0.X.Upsilon.C},
which generates the transformation
\begin{equation}\label{ex0.dyn.C.group}
  C\to C^\dagger = C -\varepsilon . 
\end{equation}
The resulting flow $(q,\dot{q})\to (q^\dagger,\dot{q}^\dagger)$ is given by
\begin{equation}\label{ex0.dyn.q.flow}
  \frac{d q^\dagger}{d\varepsilon} = P_{(\Upsilon)}(t,q^\dagger,C^\dagger)
\end{equation}
with $\dot{q}^\dagger$ subsequently obtained from 
the transformation \eqref{ex0.dyn.C.group}
combined with the expression \eqref{ex0.C} for $C$.
Unfortunately,
once the dynamical symmetry components \eqref{ex0.P.tp1}--\eqref{ex0.P.ip}
are substituted into the flow equation \eqref{ex0.dyn.q.flow},
this yields a differential equation that cannot be solved explicitly.

Instead, consider the equivalent flow given by the vector field \eqref{Y.gen}, 
$\Y_{(\Upsilon)} =  \tau\Dt + \X_{(\Upsilon)}^\solnsp$,
with $\tau(t,q,C)$ being a gauge function
which will be chosen to simplify the components of this vector field.
Here both terms on the right-hand side are expressed as vector fields
in the space $(t,q,C)$:
\begin{equation}
  \Dt = \partial_t  + \sigma(t)^{-1}\big(  \sigma'(t) q   \pm \sqrt{2Cq^2 +\beta  \sigma(t)^2}\, q^{-1} \big)\partial_q 
\end{equation}
since $\Dt C=0$;
and
\begin{equation}
  \X^\solnsp_{(\Upsilon)} = P_{(\Upsilon)} \partial_q  -\partial_C
\end{equation}
via the symmetry action \eqref{ex0.X.Upsilon.C}. 
Thus, 
\begin{equation}\label{ex0.Y.Upsilon.gauge}
  \Y_{(\Upsilon)} =
  \tau\partial_t 
  + \big( P_{(\Upsilon)} + \tau \sigma(t)^{-1}( \sigma'(t) q   \pm \sqrt{2C q^2 +\beta  \sigma(t)^2 }\, q^{-1})  \big)\partial_q
  -\partial_C . 
\end{equation}
This vector field has the same action as $\X^\solnsp_{(\Upsilon)}$ does
on any locally conserved integral. 
A useful gauge choice is 
\begin{equation}
  \tau = -\lambda\frac{\sigma(t)}{C\sigma'(t)}
\end{equation}
where $\lambda$ is a constant to be determined, 
which gives
\begin{equation}\label{ex0.Y.Upsilon}
  \Y_{(\Upsilon)} =
  -\lambda\frac{\sigma(t)}{C\sigma'(t)} \partial_t
  + \Big( P_{(\Upsilon)}(t,q,C) -\lambda\frac{q}{C}
  \mp \frac{|\sigma(t)|}{2C\sigma'(t) q} \sqrt{2C q^2 +\beta  \sigma(t)^2} 
  \Big)\partial_q
  -\partial_C . 
\end{equation}
The flow resulting from this vector field \eqref{ex0.Y.Upsilon} on $(t,q,C)$ 
is given by
\begin{subequations}
\begin{gather}
  \frac{dt^\dagger}{d\varepsilon}
  = -\lambda\frac{\sigma(t^\dagger)}{C^\dagger\sigma'(t^\dagger)} , 
  \label{ex0.dyn.t.ode}
  \\
  \frac{dq^\dagger}{d\varepsilon}
  = P_{(\Upsilon)}(t^\dagger,q^\dagger,C^\dagger)
  -\lambda\frac{q^\dagger}{C^\dagger}
  \mp \frac{|\sigma(t)|}{2C^\dagger\sigma'(t^\dagger)q^\dagger}\sqrt{2C^\dagger q^\dagger{}^2+\beta  \sigma(t^\dagger)^2} , 
    \label{ex0.dyn.q.ode}
\end{gather}
\end{subequations}
together with the transformation \eqref{ex0.dyn.C.group},
where $\varepsilon=0$ represents the identity.
Observe that the differential equation \eqref{ex0.dyn.t.ode} is separable.
To simplify the differential equation \eqref{ex0.dyn.q.ode},
make a change of variables
$\varepsilon$ to $C$, and $q$ to $y= C\sqrt{2Cq^2+\beta  \sigma(t)^2}$.
Then the choice $\lambda =-1$ leads to a linear differential equation 
in each of the cases \eqref{ex0.P.tp1}--\eqref{ex0.P.ip} for $P_{(\Upsilon)}$:
\begin{subequations}\label{ex0.y.ode}
\begin{equation}
\frac{d y^\dagger}{dC^\dagger}
=
\sqrt{C^\dagger{}^2 (\sigma(t^\dagger)\sigma'(t^\dagger)){}^{-2} +\beta}^{-1}
-\beta 2 C^\dagger \sigma(t^\dagger) |\sigma'(t^\dagger)|^{-1}
\end{equation}
in the turning-point cases \eqref{ex0.P.tp1}--\eqref{ex0.P.tp2};
\begin{equation}
\frac{d y^\dagger}{dC^\dagger}
= 2 C^\dagger \sigma(t^\dagger) |\sigma'(t^\dagger)|^{-1}
+\big( C^\dagger{}^2 \omega(t^\dagger)^{-1} -\sigma(t^\dagger)^2 \big)
\sqrt{2C^\dagger{}^2 \omega(t^\dagger)^{-1} -\sigma(t^\dagger)^2}^{-1}
\end{equation}
in the inertial-point case \eqref{ex0.P.ip}.
\end{subequations}
In all cases, the differential equation \eqref{ex0.dyn.t.ode} for $t^\dagger$
has the solution
\begin{equation}\label{ex0.dyn.t}
  \sigma(t^\dagger) C^\dagger = \sigma C . 
\end{equation}
This enables obtaining the quadrature of the preceding differential equations \eqref{ex0.y.ode}
via a change of variable $C^\dagger$ to $t^\dagger$,
yielding 
\begin{subequations}\label{ex0.y.group}
\begin{equation}
  y^\dagger =
  -\sgn(\sigma'(t))  C\sigma(t) \int_t^{t^\dagger} \Big(
  \frac{2\sigma(t)}{\sigma(x)^2}
  +  \frac{\sigma'(x)^2}{\sqrt{C^2 \sigma(t)^2 \sigma(x)^{-2} +\beta\sigma(x)^2\sigma'(x)^2}}
    \Big) dx
\end{equation}
and
\begin{equation}
  y^\dagger =
  -\sgn(\sigma'(t)) C \sigma(t) \int_t^{t^\dagger} \Big(
  \frac{2C \sigma(t)}{\sigma(x)^2}
  +\frac{|\sigma'(x)|\big(C\sigma(t) \omega(x)^{-1}\sigma(x)^{-2} -\sigma(x)\big)}{\sqrt{\big(2C\sigma(t) \omega(x)^{-1} -\sigma(x)^3\big)\sigma(x)}}
  \Big) dx , 
\end{equation}
\end{subequations}
respectively. 
Inverting the changes of variable and using the transformation \eqref{ex0.dyn.C.group}
gives 
\begin{equation}\label{ex0.dyn.t.q}
  t^\dagger  = \sigma^{-1}((1-\varepsilon) \sigma(t)),
    \quad
  q^\dagger
  =\sgn(q) \sqrt{1-\varepsilon}^{-3} C^{-3} \sqrt{\tfrac{1}{2}(y^\dagger{}^2 - \beta  C^2 \sigma(t)^2)} . 
\end{equation}  
Finally,
expression \eqref{ex0.dotq} for $\dot{q}$ 
combined with the transformation \eqref{ex0.dyn.C.group}
gives 
\begin{equation}\label{ex0.dyn.dotq}
  \dot{q}^\dagger =  \sigma(t^\dagger)^{-1}\big(
  \sigma'(t^\dagger) q^\dagger   \pm \sqrt{2(1-\varepsilon)C +\beta  \sigma(t^\dagger)^2 q^\dagger{}^{-2}}
  \big) . 
\end{equation}

Hence, the transformations \eqref{ex0.y.group}--\eqref{ex0.dyn.dotq}
represent the explicit Lie group of dynamical symmetry transformations 
arising from each of the locally conserved integrals \eqref{ex0.betais1.Upsilon.tp}--\eqref{ex0.betais-1.Upsilon.ip}.

\section{Geodesics of spheroids}\label{sec:spheroid}\label{sec:ex.1}

A spheroid in $\Rnum^3$ is a surface of revolution of an ellipse 
with principal axes $a$ and $b$, 
which is given by 
\begin{equation}
  (x/b)^2 + (y/b)^2 + (z/a)^2 = 1,
  \quad
  a\neq b . 
\end{equation}
in Cartesian coordinates $(x,y,z)$.
The spheroid is called oblate if $b>a$, or prolate if $a>b$.
Its eccentricity is defined to be $e= \sqrt{|1- R^2|}$
where $R = a/b$ is called the aspect ratio.

The intrinsic geometry of a spheroid is determined by the line element
\begin{equation}
ds^2 = \sin(\theta)^2 d\phi^2 + (\cos(\theta)^2 + R^2 \sin(\theta)^2) d\theta^2
\end{equation}
using spheroidal coordinates with respect to the axis of symmetry,
where $\theta\in[0,\pi]$ is the azimuthal angle and $\phi\in[0,2\pi)$ is polar angle. 
Geodesics are described by the equations
\begin{equation}\label{ex1.eom}
\ddot{\phi} = - 2\cot(\theta)\, \dot{\phi}\, \dot{\theta}, 
\quad
\ddot{\theta} = \frac{\sin(\theta)\cos(\theta)( \dot{\phi}^2 + (1-R^2)\dot{\theta}^2 )}{\cos(\theta)^2 + R^2\sin(\theta)^2} , 
\end{equation}
which arise from the Lagrangian
\begin{equation}\label{ex1.Lagr}
\Lagr = \tfrac{1}{2}\big(
\sin(\theta)^2 \dot{\phi}^2 + (\cos(\theta)^2 + R^2 \sin(\theta)^2) \dot{\theta}^2
\big)
\end{equation} 
where $t$ denotes an affine parameter.
The Hamilton variables are
$q^i=(\theta,\phi)$
and $p_i =\partial_{\dot{q}^i}\Lagr = ((\cos(\theta)^2 + R^2 \sin(\theta)^2 \dot{\theta}, \sin(\theta)^2\dot{\phi})$.

\subsection{Point symmetries}

The manifest point symmetries of the Lagrangian \eqref{ex1.Lagr} consist of
the polar rotation symmetry
\begin{equation}\label{ex1.rotation}
  \X_\text{rot.}=\partial_\phi , 
\end{equation}
and the affine-translation symmetry
\begin{equation}
  \X_\text{trans.} =\partial_t
\end{equation}
whose equivalent formulation \eqref{point.symm} is given by 
\begin{equation}\label{ex1.translation}
  \X_\text{trans.}   = - (\dot{\phi}\partial_\phi + \dot{\theta}\partial_\theta) . 
\end{equation}
Solving the determining condition \eqref{inv.cond}
shows that no further variational point symmetries \eqref{point.symm} exist.

\subsection{Conserved integrals and Liouville integrability}

Through Noether's theorem (cf Theorem~\ref{thm:noether}),
the variational point symmetries \eqref{ex1.rotation} and \eqref{ex1.translation}
yield constants of motion 
\begin{equation}
  C_\text{rot.} = \parder{\Lagr}{\dot{\phi}} 
\end{equation}
and
\begin{equation}
C_\text{trans.} = -\dot{\phi} \parder{\Lagr}{\dot{\phi}}-\dot{\theta} \parder{\Lagr}{\dot{\theta}} + \Lagr 
= -\Lagr . 
\end{equation}
They respectively describe 
angular momentum $C_\text{rot.}=L$ and energy $C_\text{trans.} = -E$: 
\begin{equation}\label{ex1.L.E}
  L = \sin(\theta)^2 \dot{\phi},
  \quad
  E= \tfrac{1}{2}\big( \sin(\theta)^2 \dot{\phi}^2 + (\cos(\theta)^2 + R^2 \sin(\theta)^2) \dot{\theta}^2 \big) . 
\end{equation}

The energy yields a Hamiltonian \eqref{Ham.gen} for the geodesic equations,
$\Ham = E=-C_\text{trans.}$.
Consequently, conservation of angular momentum is equivalent to the Poisson bracket
$\{C_\text{rot.},C_\text{trans.}\}=0$. 
The geodesic equations thereby possess two commuting constants of motion,
and hence they constitute a locally Liouville-integrable system. 

To obtain action-angle variables \eqref{action.angle.variables},
evaluate the integral expression \eqref{S.action.angle} given by 
\begin{equation}
  \S = \int \parder{\Lagr}{\dot{\phi}}(\phi,\theta,E,L)\, d\phi + \parder{\Lagr}{\dot{\theta}}(\phi,\theta,E,L)\, d\theta , 
\end{equation}
with the use of
\begin{equation}
  \dot{\phi} =  L/\sin(\theta)^2,
  \quad
  \dot{\theta} =\sgn(\dot{\theta}) \sqrt{(2E- L^2/\sin(\theta)^2)}/\sqrt{\cos(\theta)^2 + R^2 \sin(\theta)^2} . 
  \end{equation}
This yields
\begin{equation}
  \S =  L \phi + \sgn(\dot{\theta}) \int  \sqrt{\cot(\theta)^2 + R^2} \sqrt{2E \sin(\theta)^2 - L^2}\, d\theta
\end{equation}
from which the action-angle variables arise 
by taking derivatives with respect to $L$ and $E$. 
Let
\begin{equation}\label{ex1.angle.L}
  \Theta:= \parder{\S}{L}
  =   \phi -   s\int \sqrt{\frac{\cot(\theta)^2 + R^2}{C \sin(\theta)^2 -1}}\, d\theta
\end{equation}
and
\begin{equation}\label{ex1.angle.E}
  T:= t - \parder{\S}{E}
  = t - s L^{-1} \int \sqrt{\frac{\cos(\theta)^2 + R^2\sin(\theta)^2}{C\sin(\theta)^2 -1}}\sin(\theta)\, d\theta
\end{equation}
where $C= 2E/L^2 \geq 1$ and $s=\sgn(\dot{\theta})\sgn(L)$. 
These expressions \eqref{ex1.angle.L} and \eqref{ex1.angle.E}
will represent locally conserved integrals
provided that the integration constant is chosen
using the dynamical features of the geodesic motion. 
A natural choice consists of a turning point  where $\dot{\theta}=0$.
The constants of motion \eqref{ex1.L.E} can be combined to find that
there are two turning points given by 
$\sin(\theta_*) = |L|/(\sqrt{2E}) = 1/\sqrt{C}$,
yielding 
\begin{equation} 
  \theta_* = \arcsin(1/\sqrt{C}), \pi - \arcsin(1/\sqrt{C})
\end{equation}
with $0<\arcsin(1/\sqrt{C})\leq \tfrac{1}{2}\pi$. 
The geodesic equations \eqref{ex1.eom} shows that
the first turning point is the angle at which $\theta(t)$ is a minimum,
which will be used hereafter, 
while the second turning point is where $\theta(t)$ is a maximum.

With this choice,
the first action-angle variable \eqref{ex1.angle.L} 
is given by
\begin{subequations}\label{ex1.Phi}
\begin{equation}
  \Theta = \phi - s \intA(\theta;C)
\end{equation}
with
\begin{equation}\label{ex1.A}
  \intA(\theta;C) = \int_{\theta_*}^{\theta} \sqrt{\frac{\cot(\theta)^2 + R^2}{C \sin(\theta)^2  -1}}\, d\theta,
  \quad
  \theta_* = \arcsin(1/\sqrt{C}) 
\end{equation}
\end{subequations}
This is a local constant of motion, namely $\dot\Theta =0$
holds piecewise for all geodesics.
It is an analog of the angle of the \LRL/ (LRL) vector in central force dynamics
\cite{Fra,Per,Anc.Mea.Pas}. 
In particular,
geodesic curves are explicitly given by 
$\phi = \Theta + s\intA(\theta;C)$
as a function of $\theta$ in the domain $[\theta_*,\pi-\theta_*]$.

The second action-angle variable \eqref{ex1.angle.E}
yields 
\begin{subequations}\label{ex1.T}
\begin{equation}
  T = t - s L^{-1} \intB(\theta;C)
\end{equation}
with
\begin{equation}\label{ex1.B}  
  \intB(\theta;C) =   \int_{\theta_*}^{\theta} \sqrt{\frac{\cos(\theta)^2 + R^2\sin(\theta)^2}{C\sin(\theta)^2 -1}}\sin(\theta)\, d\theta,
\quad
\theta_* = \arcsin(\sqrt{C}^{-1})  . 
\end{equation}
\end{subequations}
This is a local integral of motion,
satisfying $\dot T=0$ piecewise for all geodesics. 
It is an analog of the LRL time in central force dynamics \cite{Anc.Mea.Pas}. 


Explicit expressions for $\Theta$ and $T$ are obtained by 
evaluation of the integrals \eqref{ex1.A} and \eqref{ex1.B}
in terms of elliptic integrals: 
\begin{align}
&\begin{aligned}
\intA(\theta;C) = & 
\tfrac{R^2-1}{R\sqrt{C}}
\K\big(\bar{C} \bar{R}\big)
-\sqrt{\tfrac{R^2-1}{C-1}}
\F\big(\bar{R} \cos(\theta),(\bar{C}\bar{R})^{-1}\big)
\\&\qquad
+\tfrac{1}{R\sqrt{C}}\Big(
\Pi\big(\bar{C}^2,\bar{C}\bar{R}\big)
- \Pi\big(\bar{C} \cos(\theta),\bar{C}^2,\bar{C}\bar{R}\big)
\Big) ; 
 \end{aligned}
\\
&
\intB(\theta;C) = \tfrac{R}{\sqrt{C}}\Big(
 \E\big(\bar{C}\bar{R}\big)
-\E\big(\bar{C}^{-1}\cos(\theta), (\bar{C}\bar{R})^{-1}\big)
\Big) 
\end{align}
where $\bar{C} = \sqrt{1 - C^{-1}}>0$ and $\bar{R}^2=1-R^{-2}\gtrless0$.
(Here, $\F$, $\E$, $\Pi$ denote elliptic integrals of the first, second, and third kinds,
with parameters given in Maple notation). 
The global behaviour of $\Theta$ and $T$ depends on
whether a given geodesic is closed or open.
In the case of a closed geodesic, 
$\theta$ will go from a minimum angle $\theta_*$ to a maximum angle $\pi-\theta_*$,
which describes one-half of the geodesic,
while $\phi$ changes by the angle $\intA(\pi-\theta_*;C) - \intA(\theta_*;C)$,
which must be a multiple of $\pi$.
When $\Theta$ is evaluated for a closed geodesic,
it is single-valued and yields the polar angle $\phi=\Theta$
at which the azimuthal angle $\theta=\theta_*$ is a minimum.
In contrast, for an open geodesic,
$\Theta$ will be multi-valued and yields a sequence of polar angles that differ by
\begin{equation}
\Delta\phi = \tfrac{4}{R\sqrt{C}}\Big(
(R^2-1) \K\big(\bar{C} \bar{R}\big) + \Pi\big(\bar{C}^2,\bar{C}\bar{R}\big) \Big) , 
\end{equation}
with each angle $\phi=n\Delta\phi + \Theta_0$, $n\in\Nnum^+$,
occurring when the azimuthal angle reaches its minimum $\theta=\theta_*$,
where $\Theta_0 = \Theta \mod \Delta\phi$. 
This behaviour is analogous to a precessing orbit in central force dynamics. 
For both open and closed geodesics,
the temporal quantity $T$ is always multi-valued 
and gives the value of the affine parameter $t$ at which $\phi=\Theta$ holds.
The difference in successive values of $T$ is given by
\begin{equation}
\Delta T = \tfrac{4}{R\sqrt{C}} \E\big(\bar{C} \bar{R})\big) . 
\end{equation}
In the case of a closed geodesic, $\Delta T$ describes the affine period,
namely the geodesic is traced out once for $0\leq t \leq \Delta T$.

\subsection{Variational symmetries}

Both conserved integrals $\Theta$ and $T$ are functions of $\theta$, $E$, $|L|$. 
Through the correspondence \eqref{PfromC.gen} in Noether's theorem,
they give rise to infinitesimal dynamical symmetries of the geodesic Lagrangian:
\begin{equation}\label{ex1.X.Phi}
  \X_{(\Theta)} =
 2  L^{-1} \intA_C \sqrt{C -\csc(\theta)^2} \big(
 s \sqrt{C -\csc(\theta)^2} \partial_\phi
  -   \sqrt{\cos(\theta)^2 + R^2\sin(\theta)^2}^{-1} \partial_\theta
  \big)
\end{equation}
and
\begin{equation}\label{ex1.X.T}
\begin{aligned}
  \X_{(T)} = &   L^{-2}\big(
  ( s\intB+ 2 s\intB_C (C-\csc(\theta)^2) )\partial_\phi
  \\&\qquad
-2 \intB_C \sqrt{C -\csc(\theta)^2} \sqrt{\cos(\theta)^2 + R^2\sin(\theta)^2}^{-1}\, \partial_\theta \big) .
\end{aligned}
\end{equation}
Note that the infinitesimal symmetries arising from $L$ and $E$ are respectively
\begin{gather}
  \X_{(L)} = \X_\text{rot.} = \partial_\phi,
  \label{ex1.X.L}
  \\
  \X_{(E)} = -\X_\text{trans.} = \dot{\phi}\partial_\phi +\dot{\theta}\partial_\theta ,
    \label{ex1.X.E}
\end{gather}
which are point symmetries. 

The action of these four symmetries on all of the conserved quantities
will now be computed,
from which the Poisson brackets on $L$, $E$, $\Theta$, $T$ will be derived.

Firstly,
since $E$ is the Hamiltonian (in Lagrangian variables) for the geodesic equations, 
the action of $\X^\solnsp_{(E)}$ vanishes on all constants of motion
and satisfies $\X^\solnsp_{(E)} T =-\partial_t T$ on the integral of motion.
Hence, the only non-zero action consists of 
\begin{equation}
  \X^\solnsp_{(E)} T = -1 . 
\end{equation} 
Secondly,
$\X^\solnsp_{(L)}$ acts geometrically as a polar rotation,
which immediately implies 
$\X^\solnsp_{(L)} L = 0$, $\X^\solnsp_{(L)} E = 0$, $\X^\solnsp_{(L)} T = 0$,
and
\begin{equation}
  \X^\solnsp_{(L)} \Theta = 1 
\end{equation}
since $\Theta$ represents a polar angle. 

Next, the action of $\pr\X_{(\Theta)}$ on $L$, $E$, and $\Theta$ is determined
from the preceding results by use of relation \eqref{X.PB.C1.C2}
in Theorem~\ref{thm:X.PB.C1.C2}. 
This yields
$\X^\solnsp_{(\Theta)} \Theta =0$, $\X^\solnsp_{(\Theta)} E =0$, and
\begin{equation}
  \X^\solnsp_{(\Theta)} L =-1 . 
\end{equation}  
The remaining action $\X^\solnsp_{(\Theta)} T$ is straightforward to compute explicitly by 
\begin{equation}\label{ex1.XPhi.T}
  \X^\solnsp_{(\Theta)} T  =
P_{(\Theta)}^{\theta} \partial_{\theta} T
+ \Dt P_{(\Theta)}^\theta \partial_{\dot\theta} T
+ \Dt P_{(\Theta)}^\phi \partial_{\dot\phi} T
\end{equation}
where $\partial_{\phi} T = \X_{(L)} T = \X^\solnsp_{(L)} T =0$.
The last two terms can be expressed as
$g_{\phi\phi} P_{(T)}^\phi\Dt P_{(\Theta)}^\phi$
and $g_{\theta\theta} P_{(T)}^\theta \Dt P_{(\Theta)}^\theta$,
where the components of $P_{(\Theta)}$ and $P_{(T)}$
are read off from the vector fields \eqref{ex1.X.Phi} and \eqref{ex1.X.T},
and where $g_{\theta\theta} = \cos(\theta)^2 + R^2\sin(\theta)^2$
and $g_{\phi\phi} = \sin(\theta)^2$
are the components of the Hessian matrix \eqref{g.matrix}. 
Time derivatives of $P_{(\Theta)}^\phi$ and $P_{(\Theta)}^\theta$
are easily computed via
\begin{equation}\label{ex1.Dt}
  \Dt\phi = L \csc(\theta)^2,
  \quad
  \Dt\theta = s L \sqrt{(C - \csc(\theta)^2)/(\cos(\theta)^2 + R^2 \sin(\theta)^2)} , 
\end{equation}
together with $\Dt L = 0$ and $\Dt E=0$.
The resulting terms in expression \eqref{ex1.XPhi.T} combine to yield
\begin{equation}
  \X^\solnsp_{(\Theta)} T =0 . 
\end{equation}

Finally, the action of $\X^\solnsp_{(T)}$ is obtained by again using Theorem~\ref{thm:X.PB.C1.C2},
which yields
$\X^\solnsp_{(T)} L=0$, $\X^\solnsp_{(T)}\Theta=0$, $\X^\solnsp_{(T)} T=0$,
and
\begin{equation}
  \X^\solnsp_{(T)} E = 1 .
\end{equation}

\subsection{Poisson brackets and the symmetry algebra}

The Poisson brackets of $L$, $E$, $\Theta$, $T$ follow directly from Theorem~\ref{thm:X.PB.C1.C2}:
\begin{align}
  \{E,L\} =0,
  \quad
  \{E,\Theta\} =0,
  \quad
  \{E,T\} = 1,
  \quad
  \{L,\Theta\} =-1,
  \quad
  \{L,T\} =0,
  \quad
  \{\Theta,T\} =0  .
\end{align}
Then, through Theorem~\ref{thm:C1.C2.varsymm.PB},
the corresponding variational symmetries 
form a four-dimensional abelian Lie algebra. 

The infinitesimal point symmetries \eqref{ex1.X.L} and \eqref{ex1.X.E}
generate a two-dimensional abelian Lie group $U(1)\times\Rnum$
comprising the point transformations
\begin{equation}\label{ex1.rotation.group}
  \phi^\dagger = \phi + \varepsilon \quad\text{mod $2\pi$}
\end{equation}
and 
\begin{equation}\label{ex1.translation.group}
  t^\dagger = t -\varepsilon
\end{equation}
with parameters $\varepsilon\in\Rnum$.
In contrast, the transformations generated by
the infinitesimal dynamical symmetries \eqref{ex1.X.Phi} and \eqref{ex1.X.T}
are more complicated.
They can be obtained explicitly as equivalent transformations
\eqref{Y.transformation.group} in the coordinate space $(t,\theta,\phi)$.

\subsection{Noether symmetry group}

For the infinitesimal dynamical symmetry \eqref{ex1.X.Phi},
consider the equivalent vector field \eqref{Y.gen} in the coordinate space $(t,\theta,\phi)$, 
\begin{equation}\label{ex1.Y.Phi.gauge}
  \Y_{(\Theta)} =  \tau\Dt + \X_{(\Theta)}
\end{equation}
where
\begin{equation}
  \Dt = \partial_t  + L \csc(\theta)^2 \partial_\phi + L^2 (C - \csc(\theta)^2) (\cos(\theta)^2 + R^2 \sin(\theta)^2)^{-1} \partial_\theta
\end{equation}
follows from expressions \eqref{ex1.Dt}. 
The gauge function $\tau$ will be chosen to simplify
the vector field \eqref{ex1.Y.Phi.gauge}
by imposing $\Y_{(\Theta)} \theta=0$.
This condition determines
\begin{equation}
  \tau = 2 L^{-2} \intA_C , 
\end{equation}
yielding
\begin{equation}\label{ex1.Y.Phi}
  \Y_{(\Theta)} =  2L^{-2} \intA_C\partial_t + 2L^{-1} C \intA_C\partial_\phi . 
\end{equation}
Now, the symmetry actions of $\X^\solnsp_{(\Theta)}$
combined with the property \eqref{X.C.Y.C}
directly give 
\begin{equation}\label{ex1.Phi.symm.L.E}
L^\dagger = L -\varepsilon,
\quad
E^\dagger = E, 
\end{equation}
and hence
\begin{equation}\label{ex1.Phi.symm.C}
  C^\dagger = C/(1 -\sqrt{C/(2E)}\varepsilon)^2 . 
\end{equation}
The vector field \eqref{ex1.Y.Phi} then shows that 
the symmetry action of $\Y_{(\Theta)}$ on the coordinates $(t,\theta,\phi)$ is given by 
\begin{equation}\label{ex1.Phi.dynsymm}
\theta^\dagger = \theta,
\quad
\phi^\dagger = \phi + \intA(\theta;C^\dagger) - \intA(\theta;C),
\quad
t^\dagger = t + \sqrt{2E}^{-1} \big( \sqrt{C^\dagger} \intB(\theta;C^\dagger) - \sqrt{C}\intB(\theta;C) \big) , 
\end{equation}
where the last expression has been obtained using the relation
\begin{equation}\label{ex1.A.B.rel}
  \intA_C = C \intB_C +\tfrac{1}{2}\intB . 
\end{equation}
(In particular, the contributions from the endpoint $\theta_*$ cancel out). 

For the infinitesimal dynamical symmetry \eqref{ex1.X.T},
a similar derivation leads to
\begin{equation}\label{ex1.T.symm.L.E.C}
L^\dagger = L, 
\quad
E^\dagger = E  +\varepsilon,
\quad
C^\dagger = C + 2L^{-2} \varepsilon , 
\end{equation}
and
\begin{equation}\label{ex1.T.dynsymm}
\theta^\dagger = \theta,
\quad
\phi^\dagger = \phi + \intA(\theta;C^\dagger) - \intA(\theta;C), 
\quad
t^\dagger = t + L^{-1} \big( \intB(\theta;C^\dagger) - \intB(\theta;C) \big) . 
\end{equation}

Through their dependence on $L$, $E$, and $C$,
these transformations \eqref{ex1.Phi.dynsymm} and \eqref{ex1.T.dynsymm}
represent dynamical symmetries. 

Therefore,
the Noether symmetry group of the geodesic equations
comprises the two point transformations \eqref{ex1.rotation.group}--\eqref{ex1.translation.group}
and the two dynamical transformations \eqref{ex1.Phi.dynsymm} --\eqref{ex1.T.dynsymm},
all of which are mutually commuting.

\section{Calogero-Moser-Sutherland system}\label{sec:ex.2}

In one spatial dimension,
consider three particles that interact by the potential \cite{Cal,Mos,Sut}
\begin{equation}
  V = k \big( (x_1 - x_2)^{-2} + (x_2 - x_3)^{-2} + (x_3 - x_1)^{-2} \big),
  \quad
  k>0
\end{equation}
where $x_1$, $x_2$, $x_3$ are the particles' positions,
and $k$ is the interaction coefficient. 
The equations of motion are given by 
\begin{equation}\label{ex2.eom}
\begin{aligned}
\ddot{x}_1 & = 2k \big( (x_1 - x_2)^{-3} + (x_1 - x_3)^{-3} \big),
\\
\ddot{x}_2 & = 2k \big( (x_2 - x_1)^{-3} + (x_2 - x_3)^{-3} \big),
\\
\ddot{x}_3 & = 2k \big( (x_3 - x_1)^{-3} + (x_3 - x_2)^{-3} \big) , 
\end{aligned}
\end{equation}
which come from the Lagrangian
\begin{equation}\label{ex2.Lagr}
\Lagr = \tfrac{1}{2} (\dot{x}_1{} ^2 + \dot{x}_2{}^2 + \dot{x}_3{}^2 ) - V . 
\end{equation}
The Hamiltonian variables are the same as the Lagrangian variables:
$q^i = (x_1,x_2,x_3)$ and $p_i = (\dot{x}_1,\dot{x}_2,\dot{x}_3)$.

\subsection{Point symmetries}

The Lagrangian \eqref{ex2.Lagr} is manifestly invariant under
a translation of the center of mass $\bar{x}=\tfrac{1}{3}(x_1+x_2+x_3)$ 
consisting of 
\begin{equation}\label{ex2.com}
  \X_\text{c.o.m.}=\partial_{x_1} + \partial_{x_2} + \partial_{x_3} , 
\end{equation}
and a time translation
\begin{equation}
  \X_\text{trans.} =-\partial_t
\end{equation}
with the equivalent formulation \eqref{point.symm} given by 
\begin{equation}\label{ex2.translation}
  \X_\text{trans.}   = \dot{x_1}\partial_{x_1}+\dot{x_2}\partial_{x_2}+\dot{x_3}\partial_{x_3} .
\end{equation}
These point symmetries respectively act as
\begin{equation}\label{ex2.com.transformation}
  x_i{}^\dagger = x_i + \varepsilon,
  \quad
  i=1,2,3 
\end{equation}
and
\begin{equation}\label{ex2.time.transformation}
  t^\dagger = t - \varepsilon
\end{equation}
with parameter $\varepsilon\in\Rnum$. 

There are additional invariances of the Lagrangian,
which are obtained by solving the determining condition \eqref{inv.cond}
for variational point symmetries \eqref{Y.point.symm}.
They consist of a Galilean boost of the center of mass 
\begin{equation}\label{ex2.Galilean}
  \X_\text{Gal.}   = t(\partial_{x_1} + \partial_{x_2} + \partial_{x_3}) , 
\end{equation}
which acts as 
\begin{equation}\label{ex2.Galilean.transformation}
  x_i{}^\dagger = x_i +t\varepsilon,
  \quad
  i=1,2,3 ; 
\end{equation}
a dilation
\begin{equation}\label{ex2.dilation}
 \X_\text{dil.}   = -2t\partial_t - x_1\partial_{x_1} - x_2\partial_{x_2} - x_3\partial_{x_3} , 
\end{equation}
which acts as
\begin{equation}
  t^\dagger = t e^{-2\varepsilon}, 
  \quad
  x_i{}^\dagger = x_i e^{-\varepsilon},
  \quad
    i=1,2,3 ; 
\end{equation}
and a conformal (non-rigid) scaling 
\begin{equation}\label{ex2.conformal}
  \X_\text{conf.}   = -t^2\partial_t - t(x_1\partial_{x_1} + x_2\partial_{x_2} + x_3\partial_{x_3} ), 
\end{equation}  
acting as
\begin{equation}
  t^\dagger = t/(1 +t\varepsilon),
  \quad
  x_i{}^\dagger = x_i/(1 +t\varepsilon),
  \quad
    i=1,2,3 .
\end{equation}  
The equivalent form \eqref{point.symm} for the latter two point symmetries is
respectively given by 
\begin{gather}
  \X_\text{dil.}   = (2t\dot{x}_1 -x_1)\partial_{x_1} + (2t\dot{x}_2 -x_2)\partial_{x_2} + (2t\dot{x}_3 -x_3)\partial_{x_3} , 
  \\
  \X_\text{conf.}   = (t^2\dot{x}_1 -tx_1)\partial_{x_1} + (t^2\dot{x}_2 -tx_2)\partial_{x_2} + (t^2\dot{x}_3 -tx_3)\partial_{x_3} . 
\end{gather}

These five variational point symmetries form a Lie algebra with the commutator structure
\begin{subequations}\label{ex2.varsymms.commutators}
\begin{gather}
[\X_\text{c.o.m.}, \X_\text{trans.}]=0, 
\quad
[\X_\text{c.o.m.}, \X_\text{Gal.}]=0,
\quad
[\X_\text{Gal.}, \X_\text{conf.}]= 0,
\label{ex2.commuting.varsymms}
\\
[\X_\text{dil.},\X_\text{c.o.m.}]=\X_\text{c.o.m.},
\quad
[\X_\text{dil.},\X_\text{trans.}]=2\X_\text{trans.},
\\
[\X_\text{dil.},\X_\text{Gal.}]=-\X_\text{Gal.},
\quad
[\X_\text{dil.}, \X_\text{conf.}]=-2\X_\text{conf.},    
\\
[\X_\text{Gal.},\X_\text{trans.}]=\X_\text{c.o.m.},
\quad
[\X_\text{c.o.m.}, \X_\text{conf.}]=-\X_\text{Gal.},
\quad
[\X_\text{trans.}, \X_\text{conf.}]=-2\X_\text{dil.} .
\end{gather}
\end{subequations}

\subsection{Conserved quantities and Poisson brackets}

Through Noether's theorem (cf Theorem~\ref{thm:noether}),
the five variational symmetries \eqref{ex2.com}, \eqref{ex2.translation}, \eqref{ex2.Galilean}, \eqref{ex2.dilation}, \eqref{ex2.conformal} 
yield five locally conserved integrals:\\
total momentum
\begin{equation}\label{ex2.mom}
P= \dot{x}_1 + \dot{x}_2 + \dot{x}_3 ; 
\end{equation}
energy
\begin{equation}\label{ex2.ener}
E = \tfrac{1}{2} (\dot{x}_1{}^2 + \dot{x}_2{}^2 + \dot{x}_3{}^2) + V ; 
\end{equation}
Galilean momentum
\begin{equation}\label{ex2.Galmom}
K = tP -(x_1+x_2+x_3) ;
\end{equation}
dilational energy
\begin{equation}\label{ex2.dil.ener}
  E_\text{dil.} = tE  -\tfrac{1}{2}(\dot{x}_1 x_1+\dot{x}_2 x_2+\dot{x}_3 x_3) ;
\end{equation}
conformal energy
\begin{equation}\label{ex2.conf.ener}
  E_\text{conf.} = t^2 E - t(\dot{x}_1 x_1+\dot{x}_2 x_2+\dot{x}_3 x_3)
   + \tfrac{1}{2}(x_1{}^2+x_2{}^2+x_3{}^2) .
\end{equation}
The first two are constants of motion, while the remaining three are integrals of motion.
Since the dynamical system \eqref{ex2.eom} is autonomous,
two constants of motion can be formed from the integrals of motion
by the method explained in \Ref{Anc2026}.
Specifically, they are obtained as solutions of the linear homogeneous PDE
\begin{equation}
  P \partial_K F + E \partial_{E_\text{dil.}} F + E_\text{dil.} \partial_{E_\text{conf.}} F =0
\end{equation}
for $F(K,E_\text{dil.},E_\text{conf.})$. 
This yields
\begin{gather}
C_1 = P E_\text{dil.} -K E 
= (x_1+x_2+x_3) E - \tfrac{1}{2}(\dot{x}_1 + \dot{x}_2 + \dot{x}_3)(\dot{x}_1 x_1+\dot{x}_2 x_2+\dot{x}_3 x_3), 
\label{ex2.C1}\\
C_2 = E E_\text{conf.} - E_\text{dil.}^2
= \tfrac{1}{2}(x_1{}^2+x_2{}^2+x_3{}^2)E 
 - \tfrac{1}{4}(\dot{x}_1 x_1+\dot{x}_2 x_2+\dot{x}_3 x_3)^2, 
\label{ex2.C2}
\end{gather}
which are constants of motion. 

Local Liouville integrability for this system requires having
three locally conserved integrals that are mutually commuting in the Poisson bracket.
It is straightforward to compute the Poisson brackets of
$P$, $E$, $K$, $C_1$, $C_2$ 
from relation \eqref{C1.C2.varsymm.PB} in Theorem~\ref{thm:C1.C2.varsymm.PB}.

From the commutators \eqref{ex2.varsymms.commutators}
involving $P$, $E$, $K$,
the corresponding Poisson brackets are given by
\begin{equation}
  \{P,E\} =0,
  \quad
  \{P,K\} =0,
  \quad
  \{E,K\} =P . 
\end{equation}
All of the Poisson brackets which involve $C_1$ and $C_2$
can be obtained through expressions \eqref{ex2.C1}--\eqref{ex2.C2}
via the commutators of the variational symmetries corresponding to
$P$, $E$, $K$, $E_\text{dil.}$, $E_\text{conf.}$:
\begin{gather}
\{E,C_1\} =0,
\quad
\{E,C_2\} =0,
\\
\{P,C_1\} = \tfrac{1}{2}P^2 -3E,
\quad
\{P,C_2\} = -C_1,
\quad
\{C_1,C_2\} = P C_2 , 
\\
\{K,C_1\}
= \tfrac{1}{2} P K -3 P^{-1}(KE + C_1), 
\\
\{K,C_2\}
= -( P^{-1} K C_1 + E^{-1} P C_2+ (EP)^{-1}C_1^2 ) . 
\end{gather}

A functional combination $f(P,E,C_1,C_2)$ that commutes with $P$
can be found by solving the linear homogeneous partial differential equation
\begin{equation}
  \{P,f\}=f_{C_1} \{P,C_1\} + f_{C_2}\{P,C_2\} =0 . 
\end{equation}
This yields the particular solution
\begin{equation}
  C =  (6E -P^2)C_2 - C_1^2 . 
\end{equation}
After expressions \eqref{ex2.C1}--\eqref{ex2.C2} are substituted,
this solution turns out to factorize $C = \tfrac{2}{3} E C_3$,
yielding the constant of motion
\begin{equation}\label{ex2.C3}
  C_3 =  3((x_2 - x_3)\dot{x}_1 + (x_3 - x_1)\dot{x}_2 + (x_1 - x_2)\dot{x}_3)^2
  + 12k (x_2 - x_3)(x_1 - x_2)(x_3 - x_1) \sqrt{V/k}^3 . 
\end{equation}
Since $E$ commutes with $P$,
note that $C_3$ still commutes with $P$, and it also commutes with $E$.

The variational point symmetries have thus led to
a set of five functionally independent conserved integrals,
comprising one integral of motion $K$
and four constants of motion $P$, $E$, $C_3$, $C_1$ or $C_2$,
of which the first three are mutually commuting. 
As a consequence, the dynamical system \eqref{ex2.eom} is locally Liouville integrable.
In fact, all of the constants of motion are globally conserved
because they are polynomial in the dynamical variables,
so the integrability holds globally.

\subsection{Liouville integrability} 

Action-angle variables \eqref{action.angle.variables} arise from
the integral expression \eqref{S.action.angle}
\begin{equation}
  \S = \int  \parder{\Lagr}{\dot{x}^i}(x,P,E,C_3) \, dx^i
\end{equation}
with $\dot{x}^i$, $i=1,2,3$, being eliminated in terms of $P,E,C_3$. 
However, this involves solving a quartic equation
whose solution goes into the preceding integral, 
which becomes impractical to evaluate explicitly. 

An alternative approach is to take advantage of the freedom to separate
the dynamics into the motion of the center of mass 
$\bar{x} = \tfrac{1}{3}(x_1 + x_2 + x_3)$ 
and the relative motions among the particles given by the displacements 
$y_1 = x_2 - x_3$, $y_2 = x_3 - x_1$, $y_3 = x_1 - x_2$
which satisfy $y_1+ y_2 + y_3 =0$.
Since total momentum is conserved, the center of mass moves with constant speed,
$\dot{\bar{x}} = \tfrac{1}{3} P$.
The relative motions then look simplest in the rest frame of the center of mass \cite{Mar},
where $P=0$.
There will be two degrees of freedom which can be taken as
two linearly independent combinations of the relative displacements $y_1$, $y_2$, $y_3$.
These degrees of freedom can be chosen
so that the Hamiltonian $H=E$ becomes diagonal. 
Without loss of generality, let $y=y_3+ ay_1$ and $z=y_1+b y_3$, 
where $a$ and $b$ are constants to be determined.
Changing variables from $(x_1,x_2,x_3)$ to $(\bar{x},y,z)$ in the Hamiltonian
and imposing the diagonality condition yields $ab -2(a+b) +1 =0$.
A solution that is both simple and useful for subsequent computations is
$a=\tfrac{1}{2}$, $b=0$.

The diagonal Hamiltonian in the center-of-mass rest frame is thereby
(up to a scaling factor) 
\begin{equation}\label{ex2.tilE}
  \tilde E = \tfrac{1}{2} \dot{y}^2 + \tfrac{3}{8} \dot{z}^2
  + \tfrac{3}{2} k  ( (2y)^2 + 3 z^2)^2 z^{-2} (2y +z)^{-2}(2y-z)^{-2}
\end{equation}
in terms of the relative displacements
\begin{equation}\label{ex2.yz}
  y = y_3 +\tfrac{1}{2} y_1 = x_1 - \tfrac{1}{2} x_2 -\tfrac{1}{2} x_3, 
  \quad
  z = y_1  =  x_2 - x_3 . 
\end{equation}
Similarly, the other constant of motion $C_3$ is
(up to a scaling factor) 
\begin{equation}\label{ex2.C3.polar}
  \tilde C_3 = \tfrac{3}{8} (z\dot{y} - y\dot{z})^2
  + \tfrac{3}{8} k( (2y)^2 + 3z^2)^3 z^{-2} (2y + z)^{-2} (2y - z)^{-2}  . 
\end{equation}
The associated Lagrangian is given by
\begin{equation}
  \tilde\Lagr = \tfrac{1}{2} \dot{y}^2 + \tfrac{3}{8} \dot{z}^2
  -\tfrac{3}{2} k ((2y)^2 + 3 z^2)^2 z^{-2} (2y +z)^{-2}(2y-z)^{-2} , 
\end{equation}
yielding the dynamical equations for the relative motion, 
\begin{equation}\label{ex2.yz.eom}
  \begin{aligned}
    \ddot{y} & = 96 k y ((2y)^2 + 3 z^2) (2y +z)^{-3} (2y - z)^{-3}, 
    \\
    \ddot{z} & = 4 k ((2y)^2 + 3z^2) (16 y^4 -24 y^2 z^2 -3 z^4)  z^{-3} (2y +z)^{-3} (2y - z)^{-3} . 
    \end{aligned}
\end{equation}

Now introduce polar variables adapted to $(y,z)$ \cite{Mar}:
\begin{equation}\label{ex2.polar}
  y = r\cos(\phi),
  \quad
  \tfrac{\sqrt{3}}{2} z = r\sin(\phi) . 
\end{equation}
This yields
\begin{equation}
  \tilde \Lagr = \tfrac{1}{2}( \dot{r}^2 + r^2 \dot{\phi}^2 )
  - \tfrac{1}{2}\tilde k r^{-2}   \csc(3\phi)^2 , 
\end{equation}
and 
\begin{align}
  \tilde E & = \tfrac{1}{2}( \dot{r}^2 + r^2 \dot{\phi}^2 ) + \tfrac{1}{2} \tilde k r^{-2}  \csc(3\phi)^2 , 
\label{ex2.tilE.polar}
\\
\tilde C_3 & =  \tfrac{1}{2} r^4\dot{\phi}^2 + \tfrac{1}{2} \tilde k  \csc(3\phi)^2 ,
\label{ex2.tilC3.polar}
\end{align}
where $\tilde k =(\tfrac{9}{2})^2 k$.
For this polar form of the Lagrangian and commuting constants of motion,
action-angle variables can be derived straightforwardly from expression \eqref{S.action.angle}:
\begin{equation}
  \S = \int \parder{\tilde\Lagr}{\dot{r}}(r,\phi,\tilde E,\tilde C_3) \, dr + \parder{\tilde\Lagr}{\dot{\phi}}(r,\phi,\tilde E,\tilde C_3) \, d\phi
\end{equation}
with 
\begin{subequations}\label{ex2.dotphi.dotr}
\begin{align}
  \dot{\phi} & = \sgn(\dot{\phi}) r^{-2} \sqrt{2\tilde C_3 \sin(3\phi)^2 - \tilde k} , 
  \quad
  \sin(3\phi)^2 \geq \tilde k/(2\tilde C_3),
  \label{ex2.dotphi}
  \\
  \dot{r} & = \sgn(\dot{r}) \sqrt{2(\tilde E - \tilde C_3 r^{-2})},
  \quad
  r^2\geq \tilde C_3/\tilde E . 
  \label{ex2.dotr}
\end{align}
\end{subequations}
The integral has the form 
\begin{equation}
  \S = \int \sgn(\dot{r}) \sqrt{2(\tilde E -\tilde C_3 r^{-2})}\, dr + \sgn(\dot{\phi}) \sqrt{2\tilde C_3 - \tilde k\csc(3\phi)^3}\, d\phi 
\end{equation}
from which the action-angles arise by
taking derivatives with respect to $\tilde E$ and $\tilde C_3$.
Thus, 
\begin{equation}\label{ex2.angle.E}
  T:=  t - \parder{\S}{\tilde E}
  = t - \int \sgn(\dot{r}) \sqrt{2(\tilde E -\tilde C_3 r^{-2})}^{-1}\, dr 
\end{equation}
and
\begin{equation}\label{ex2.angle.C3}
  \Theta: = \parder{\S}{\tilde C_3}
  = -\int \sgn(\dot{r}) r^{-2} \sqrt{2(\tilde E -\tilde C_3 r^{-2})}^{-1}\, dr
  + \int \sgn(\dot{\phi}) \sqrt{2\tilde C_3 - \tilde k\csc(3\phi)^3}^{-1}\, d\phi . 
\end{equation}

These expressions \eqref{ex2.angle.E} and \eqref{ex2.angle.C3}
for the action-angle variables 
will represent locally conserved integrals
once a choice of integration constant is made 
using the intrinsic dynamical features of particles' interaction.
Inequality \eqref{ex2.dotr} shows that $r$ has a minimum
$r_*=\sqrt{\tilde C_3/\tilde E}$
where $\dot{r}=0$, which is a turning point.
This choice for the integration constant in expression \eqref{ex2.angle.E}
gives
\begin{equation}\label{ex2.T.polar}
T  = t - \sgn(\dot{r}) \tilde E^{-1} \sqrt{\tfrac{1}{2}(\tilde E r^2 -\tilde C_3)} . 
\end{equation}
The same choice is required for the $r$-integral in expression \eqref{ex2.angle.C3},
while a separate choice is needed for the $\phi$-integral.
A turning point can again be used, given by $\dot{\phi}=0$,
where $\phi$ has a minimum or maximum.
The choice of the minimum angle
$\phi_*= \tfrac{1}{3}\arcsin\big(\sqrt{\tilde k/(2\tilde C_3)}\big)$
determined from inequality \eqref{ex2.dotphi}
yields 
\begin{equation}\label{ex2.Psi.polar}
  \Psi =
  \sgn(\dot{\phi}) \arctan\Big(\sec(3\phi) \sqrt{\sin(3\phi)^2 - \tilde k/(2\tilde C_3)}\Big)
-\sgn(\dot{r}) 3\arctan\Big(\sqrt{(\tilde E/\tilde C_3) r^2 - 1}\Big)
\end{equation}
where expression \eqref{ex2.angle.C3} has been scaled
to absorb a constant factor $\sqrt{2\tilde C_3}^{-1} =\Theta/\Psi$.

The resulting expressions \eqref{ex2.T.polar} and \eqref{ex2.Psi.polar}
are locally conserved integrals, 
namely $\dot{T} =0$ and $\dot{\Psi}=0$ hold piecewise for all trajectories. 
Their physical meaning can be understood in terms of the relative displacements
\begin{equation}
y_1 = \tfrac{2}{\sqrt{3}} r \sin(\phi),
\quad
y_2= \tfrac{2}{\sqrt{3}} r \sin(\phi + \tfrac{2}{3}\pi),
\quad
y_3= \tfrac{2}{\sqrt{3}} r \sin(\phi + \tfrac{4}{3}\pi) , 
\end{equation}
satisfying $y_1+y_2+y_3=0$.
From the constants of motion \eqref{ex2.tilE.polar} and \eqref{ex2.tilC3.polar},
the relative motion of the particles is described by
the effective potentials
$V^r = r^{-2} \tilde C_3$ and $V^\phi = \tfrac{1}{2}\tilde k \csc(3\phi)^2$,
which show $\phi$ will oscillate between 
a minimum $\phi_*$ and a maximum $\tfrac{1}{3}\pi-\phi_*$,
while $r$ will either increase to infinity
or decrease to a minimum $r_*$ and then increase. 
The time at which $r$ reaches its minimum is given by $T$.
This quantity is globally continuous, namely $T$ is a global integral of motion.
When $r$ is a minimum, $\phi$ is determined by the value of $\Psi$ through the relation
$\tan(|\Psi|) = \sec(3\phi|_{r_*}) \sqrt{\sin(3\phi|_{r_*})^2 - \tilde k/(2\tilde C_3)}$.
As a consequence, $\Psi$ is a globally continuous constant of motion. 
Note that $\Psi$ combined with the polar variables \eqref{ex2.polar}
also determines the shape of the relative-motion trajectories
$(y(t),z(t))$ in the $yz$-plane:
\begin{subequations}\label{ex2.trajectory}
\begin{equation}
  \sgn(\alpha)\cos(3\phi) = \sgn(\dot{\phi})\sqrt{1-\tilde k/(2\tilde C_3)}\, \sin(\sgn(\dot{r})\alpha -\Psi)
\end{equation}
where
\begin{equation}
  \cot(\alpha) = F(3-F^2) (1-3F^2)^{-1},
  \quad
  F = \sqrt{(\tilde E/\tilde C_3) r^2 -1} . 
\end{equation}
\end{subequations}

The system \eqref{ex2.yz.eom} for the relative motion of the particles
thus possesses the two commuting constants of motion $\tilde E$ and $\tilde C_3$,
the additional constant of motion $\Psi$,
and the integral of motion $T$.
All of these conserved quantities are globally continuous,
and hence this system is Liouville integrable. 

Both the integral of motion \eqref{ex2.T.polar}
and the constant of motion \eqref{ex2.Psi.polar} 
correspond to conserved quantities in the original system \eqref{ex2.eom}
through inverting the change of variables \eqref{ex2.yz} and \eqref{ex2.polar},
\begin{equation}
  r = \tfrac{1}{\sqrt{2}}\sqrt{(x_1 - x_2)^2 + (x_2 - x_3)^2 + (x_3-x_1)^2},
  \quad
  \tan(3\phi) = \tfrac{1}{\sqrt{3}}\frac{(x_1-x_2)(x_2-x_3)(x_3-x_1)}{(x_1-\bar{x})(x_2-\bar{x})(x_3-\bar{x})},
\end{equation}
and using the relations
\begin{equation}\label{ex2.scal.rels}
  \tilde E =\tfrac{3}{2} E -\tfrac{1}{4} P^2,
  \quad
  \tilde C_3 = \tfrac{1}{8} C_3 . 
\end{equation}
This gives 
\begin{equation}\label{ex2.T}
  T = t + (6E -P^2)^{-1} \big( (x_1+x_2+x_3)P -3(x_1\dot{x}_1 +x_2\dot{x}_2 +x_3\dot{x}_3) \big)
\end{equation}
and
\begin{equation}\label{ex2.Psi}
  \begin{aligned}
  \Psi = & 
  -\arctan\Big(\frac{3(x_1-x_2)(x_2-x_3)(x_3-x_1)( (x_2-x_3)\dot{x}_1 + (x_3 -x_1)\dot{x}_2 + (x_1 -x_3)\dot{x}_3 )}{(\bar{x} -x_1)(\bar{x} -x_2)(\bar{x} -x_3)\sqrt{C_3}}\Big)
  \\&\qquad
  -3\arctan\Big(\frac{3(x_1\dot{x}_1 + x_2\dot{x}_2 +x_3\dot{x}_3) -(x_1+x_2+x_3)P}{\sqrt{C_3}}\Big) . 
  \end{aligned}
\end{equation}
The latter quantity has an alternative formulation
in terms of a constant of motion 
\begin{equation}\label{ex2.cubc}
C_4 = \tfrac{1}{3}(\dot{x}_1{}^3 + \dot{x}_2{}^3 + \dot{x}_3{}^3)
+ (\dot{x}_1 +\dot{x}_2) V_3
+ (\dot{x}_2 +\dot{x}_3) V_1 
+ (\dot{x}_3 +\dot{x}_1) V_2 
\end{equation}
which is polynomial in velocities, 
where
\begin{equation}
  V_1 = k (x_2 - x_3)^{-2},
  \quad
  V_2 = k (x_3 - x_1)^{-2},
  \quad
  V_3 = k (x_1 - x_2)^{-2} . 
\end{equation}
This arises from a matrix Lax pair $\partial_t L = [L,A]$ 
that exists for the system \eqref{ex2.eom} \cite{Mos}. 
Taking the trace of powers of $L$ yields
constants of motion given by polynomials in the velocities,
specifically the total momentum $P$, energy $E$, and $C_4$.
(All other trace-powers are polynomials of those three.)
When specialized to the rest frame of center of mass 
and expressed in polar variables,
$C_4$ yields an algebraic equation for $\cos(3\phi)$ in terms of $r$,
which is equivalent to the trajectory equation \eqref{ex2.trajectory},
where 
\begin{equation}\label{ex2.Psi.C4}
\tan(\Psi)
= \frac{\sgn(\dot{\phi})\sgn(\dot{r}) \tilde C_4}{\sqrt{(1-\tilde k/(2\tilde C_3))(2\tilde E)^3 - \tilde C_4{}^2}},
\quad
\tilde C_4 = \tfrac{27}{2} C_4 - 9 PE + P^3 . 
\end{equation}


Altogether,
the system possesses five constants of motion $P$, $E$, $C_1$, $C_3$, $\Psi$ or $C_4$,
and two integrals of motion $K$, $T$. 
Note that $K$ and $T$ can be combined to yield 
\begin{equation}\label{ex2.extra}
  PT - K
  = 3(6E -P^2)^{-1} C_1 . 
\end{equation}
Thus, there is a total of six functionally-independent conserved integrals,
in accordance with the three degrees of freedom of the system \eqref{ex2.eom}.

\subsection{Dynamical symmetries}

The conserved integrals that generate
the Noether symmetry group of the system \eqref{ex2.eom} 
are given by $P$, $E$, $C_3$, $T$, either $C_1$ or $K$, and either $C_4$ or $\Psi$. 
Since $K$ is a point symmetry, and $C_4$ is polynomial in the velocities, 
the simplest group comes from $K$, $P$, $E$, $C_3$, $C_4$,  $T$. 
To emphasize the Noether correspondence \eqref{PfromC.gen},
the notation
$\X_\text{mom.} = \X_{(P)}$,
$\X_\text{trans.} = \X_{(E)}$,
and $\X_\text{Gal.} = \X_{(K)}$
will be adopted hereafter.

Noether's theorem (cf Theorem~\ref{thm:noether}) shows that 
the constants of motion $C_3$, $C_4$, and the integral of motion $T$
each correspond to an infinitesimal dynamical symmetry:
\begin{align}
&  \X_{(C_3)} = 6 A \big( (x_2 -x_3 )\partial_{x_1} + (x_3 - x_1)\partial_{x_2} + (x_1 - x_2)\partial_{x_3} \big) ; 
\label{ex2.C3.X}
\\
 &  \X_{(C_4)}   =
  ( \dot{x}_1{}^2 +V_2 + V_3 )\partial_{x_1}
  +  ( \dot{x}_2{}^2 +V_1 + V_3 )\partial_{x_2}
  +  ( \dot{x}_3{}^2 + V_1 + V_2 )\partial_{x_3} ; 
\label{ex2.C4.X}
\\
& \X_{(\Psi)} = 
-\tfrac{2}{9}\tilde C_3^{-1} \tilde C_4^{-2} \frac{\sin(\Psi)^3}{\cos(\Psi)} 
\Big(
\tilde k (\tilde E^3/\tilde C_3) \X_{(\tilde C_3)}
+ (\tilde E^2/\tilde C_4) (\tilde k - 2\tilde C_3)
\big( 2\tilde E\X_{(\tilde C_4)} - 3\tilde C_4\X_{(\tilde E)} \big)
\Big) ; 
\label{ex2.Psi.X}
\\
&\begin{aligned}
  \X_{(T)} = & 
  6(6E-P^2)^{-2}( \bar{x} P -x_1\dot{x}_1 -x_2\dot{x}_2-x_3\dot{x}_3 ) (P \X_{(P)} - 3 \X_{(E)})
  \\&\qquad
  +   3(6E-P^2)^{-1}\big( (\bar{x}-x_1) \partial_{x_1}  + (\bar{x}-x_2) \partial_{x_2} + (\bar{x}-x_3) \partial_{x_3} \big) ; 
\end{aligned}
\label{ex2.T.X}
\end{align}
where
\begin{equation}
  A = (x_2 - x_3)\dot{x}_1 + (x_3 - x_1)\dot{x}_2 + (x_1 - x_2)\dot{x}_3 . 
\end{equation}
Note that $\X_{(\tilde C_3)}$, $\X_{(\tilde C_4)}$, $\X_{(\tilde E)}$
can be expressed in terms of $\X_{(C_3)}$, $\X_{(C_4)}$, $\X_{E)}$, and $\X_{(P)}$
through the relations \eqref{ex2.scal.rels} and \eqref{ex2.Psi.C4}. 

The action of the point and dynamical symmetries on the conserved integrals
will be useful for deriving the dynamical symmetry groups generated by
the infinitesimal symmetries \eqref{ex2.C3.X}--\eqref{ex2.T.X}
as well as for obtaining the Poisson brackets of the conserved integrals. 

Start with the three point symmetries, $\X_{(P)}$, $\X_{(E)}$, and $\X_{(K)}$. 
Since $E$ is the Hamiltonian (in Lagrangian variables), 
the action of $\X^\solnsp_{(E)}$ vanishes on all of the constants of motion, 
while it acts on the integrals of motion by 
\begin{equation}
  \X^\solnsp_{(E)} T =-1,
  \quad
  \X^\solnsp_{(E)} K = -P . 
\end{equation}
Moreover, as the velocities and the potentials $V_1$, $V_2$, $V_3$
are invariant under translations in the center of mass,
the action of $\X^\solnsp_{(P)}$ vanishes on $P$, $E$, and $C_3$, $C_4$ and $\Psi$. 
It shifts $\bar{x}$ by $1$ and $x_1\dot{x}_1 +x_2\dot{x}_2 +x_3\dot{x}_3$ by $P$,
whereby it vanishes on $T$
and acts on $K$ by 
\begin{equation}
  \X^\solnsp_{(P)} K = -3 . 
\end{equation}
The action of $\X^\solnsp_{(K)}$ leaves invariant the potentials $V_1$, $V_2$, $V_3$
while the velocities get shifted, $\X^\solnsp_{(K)} \dot{x}_i =1$, $i=1,2,3$.
This can be shown to imply that $C_3$, $T$, and $\Psi$ 
are annihilated by $\X^\solnsp_{(K)}$.
Its action on the other conserved quantities is given by 
\begin{equation}
  \X^\solnsp_{(K)} P = 3,
  \quad
  \X^\solnsp_{(K)} E = P,
  \quad
  \X^\solnsp_{(K)} C_4 = 2E . 
\end{equation}

Use of relation \eqref{X.PB.C1.C2} in Theorem~\ref{thm:X.PB.C1.C2}
applied to the previous symmetry actions directly yields
the actions of the dynamical symmetries on $P$, $E$, and $K$. 
In particular, the non-zero ones are 
\begin{equation}
  \X^\solnsp_{(T)} E  = 1,
  \quad
  \X^\solnsp_{(C_4)} K = -2E . 
\end{equation}
The remaining symmetry actions must be computed.

A straightforward albeit lengthy computation shows that
$\X^\solnsp_{(C_3)} T=0$, $\X^\solnsp_{(T)} \Psi=0$, 
and 
\begin{equation}\label{ex2.C3.X.Psi}
  \X^\solnsp_{(C_3)}\Psi = 18 \sqrt{C_3} . 
\end{equation}
The symmetry action of $\X^\solnsp_{(C_4)}$ on $C_3$, $\Psi$, and $T$
can be computed in terms of the symmetry action of $\X^\solnsp_{(\Psi)}$
through the relation \eqref{ex2.Psi.C4}.
This leads to lengthy expressions which will be omitted.

\subsection{Noether symmetry group}

Applying Theorem~\ref{thm:X.PB.C1.C2} to the symmetry actions 
directly gives the non-zero Poisson brackets of $P$, $E$, $C_3$, $\Psi$, $K$, and $T$:
\begin{gather}
 \{E,T\} =1,
  \quad
  \{P,K\} = -3,
\\
 \{E,K\} = P,
  \quad
\{\Psi,C_3\} = 18\sqrt{C_3} .
\end{gather}
Observe that the first two brackets indicate the constants of motion $E$ and $P$
are canonically conjugate to the integrals of motion $T$ and $K$ (up to scalings).
In the last bracket, 
replacement of $C_3$ by
\begin{equation}
  C_{3'}=\exp(\tfrac{1}{3}\sqrt{C_3})
\end{equation}
yields
\begin{equation}
  \{C_{3'},\Psi\} = 3C_{3'} . 
\end{equation}
Then all of the Poisson brackets of $P$, $E$, $C_{3'}$, $\Psi$, $K$, and $T$ close linearly.

As a consequence, 
Theorem~\ref{thm:C1.C2.varsymm.PB}
shows that the corresponding infinitesimal symmetries
$\X^\solnsp_{(P)}$, $\X^\solnsp_{(E)}$, $\X^\solnsp_{(C_{3'})}$, $\X^\solnsp_{(\Psi)}$, $\X^\solnsp_{(K)}$, and $\X^\solnsp_{(T)}$
form a six-dimensional Lie algebra
whose non-zero commutators are given by 
\begin{equation}\label{ex2.pointsymm.comms}
  [\X^\solnsp_{(E)},\X^\solnsp_{K}] = -\X^\solnsp_{(P)},
  \quad
  [\X^\solnsp_{(\Psi)},\X^\solnsp_{(C_{3'})}] = 3\X^\solnsp_{(C_{3'})} ,
\end{equation}
where $\X^\solnsp_{(P)}$, $\X^\solnsp_{(E)}$, $\X^\solnsp_{(C_{3'})}$, $\X^\solnsp_{(T)}$
span a four-dimension abelian subalgebra. 

The Noether symmetry group is generated by these six symmetries. 
Each symmetry itself produces 
a one-dimensional transformation group \eqref{X.transformation.group},
where the three point symmetries $\X^\solnsp_{(P)}$, $\X^\solnsp_{(E)}$, $\X^\solnsp_{(K)}$ 
give the respective point transformation groups 
\eqref{ex2.com.transformation}, 
\eqref{ex2.time.transformation}, 
\eqref{ex2.Galilean.transformation}.
The transformation groups produced by the three dynamical symmetries
$\X_{(C_{3'})}$, $\X_{(T)}$, $\X_{(\Psi)}$
will now be derived.
These transformations turn out to look simplest
in polar variables combined with the center of mass variable: $(r,\phi,\bar{x})$.

Consider, first, $\X_{(C_{3'})}=\tfrac{1}{6}(C_{3'}/\sqrt{C_3}) \X_{(C_3)}$.
A direct computation gives 
\begin{equation}\label{ex2.C3.flow}
\begin{gathered}
  \X_{(C_{3'})} \bar{x} = 0,
  \quad
  \X_{(C_{3'})} r = 0,
\\
  \X_{(C_{3'})} \phi = 2(C_{3'}/\sqrt{C_3}) r^2 \dot{\phi}
  =   \sgn(\dot{\phi})  C_{3'} \sqrt{1 -4(\tilde k/C_3)\csc(3\phi)^2} , 
\end{gathered}
\end{equation}
with use of the polar equations of motion \eqref{ex2.dotphi.dotr}
and the relations \eqref{ex2.scal.rels}.
It is easy to integrate the flow defined by the vector field \eqref{ex2.C3.flow}, 
since $\X_{(C_{3'})} C_3=0$. 
This yields the transformation group 
\begin{subequations}\label{ex2.C3.group}
\begin{gather}
  \bar{x}^\dagger =\bar{x},
  \quad
  r^\dagger =r,
\\
  \cos(3\phi^\dagger) =
  \cos(3C_{3'}\varepsilon) \cos(3\phi)
  -\sgn(\dot{\phi})\sin(3C_{3'}\varepsilon) \sqrt{\sin(3\phi)^2 - 4\tilde k/C_3}
\end{gather}
\end{subequations}
with parameter $\varepsilon\in\Rnum$. 
The explicit appearance of $C_3$ ($C_{3'}$)
indicates that these transformations are dynamical.

Next, consider $\X_{(T)}$.
Its action on the variables $(r,\phi,\bar{x})$ is given by 
\begin{equation}\label{ex2.T.flow}
  \begin{gathered}
  \X_{(T)} \bar{x} = 0,
  \quad
  \X_{(T)} r = 3 r ( 1 - 4 \dot{r}^2 (6E -P^2)^{-1} )(6E -P^2)^{-1}  ,
  \quad
  \X_{(T)} \phi = 12 r \dot{r} \dot{\phi}(6E -P^2)^{-2} . 
  \end{gathered}
\end{equation}
The flow defined by this vector field \eqref{ex2.T.flow}
is complicated to derive directly because
$\X^\solnsp_{(T)}E=1$ leads to $E^\dagger = E+\varepsilon$
whereby the ODE for $r^\dagger$ is not separable,
despite $C_3^\dagger = C_3$ and $P^\dagger=P$
due to $\X^\solnsp_{(T)}C_3=0$ and $\X^\solnsp_{(T)}P=0$.
This ODE is, however, of Bernoulli type which can be integrated explicitly.

Alternatively, it is possible to obtain an equivalent flow more simply
in the extended coordinate space $(t,r,\phi,\bar{x})$. 
This flow arises from the vector field \eqref{Y.gen},
which is given by
\begin{equation}\label{ex2.Y.T.gauge}
  \Y_{(T)} =  \tau\Dt + \X_{(T)}
\end{equation}
where $\Dt = \partial_t  + \dot{r} \partial_r + \dot{\phi} \partial_\phi  + \tfrac{1}{3}P \partial_{\bar{x}}$.
The gauge function $\tau$ here will be chosen so that $\Y_{(T)} r = 0$,
which determines
\begin{equation}
  \tau = 
  \sgn(\dot{r}) 6 (C_3 -(6E -P^2)r^2) (6E -P^2)^{-2} \sqrt{2(6E -P^2)r^2 - C_3}^{-1}
\end{equation}
after use of the polar equations of motion \eqref{ex2.dotphi.dotr}.
This yields the vector field
\begin{subequations}\label{ex2.T.Y}
\begin{gather}
 \Y_{(T)} r = 0 ,
\quad
 \Y_{(T)} \bar{x} = \tau P,
\\
\Y_{(T)} \phi =
\sgn(\dot{\phi}) \sgn(\dot{r}) \tfrac{3}{2}\sqrt{2 C_3 -8\tilde k\csc(3\phi)^2}\,
(6E^\dagger -P^2)^{-1}\sqrt{2(6E\mathstrut^\dagger -P^2)r^2 - C_3}^{-1} . 
\end{gather}
\end{subequations}
The ODE for $r^\dagger$ simply gives
\begin{subequations}\label{ex2.T.group}
\begin{equation}\label{ex2.T.Y.r}
  r^\dagger =r , 
\end{equation}  
while the ODE for $\phi^\dagger$ is separable
and the ODE for $\bar{x}^\dagger$ reduces to quadrature.
The resulting transformations are given by 
\begin{gather}
\cos(3\phi^\dagger) =
\sgn(\dot{\phi}) \big( \cos(R(r,\varepsilon)\sqrt{C_3}) \cos(3\phi) 
-\sin(R(r,\varepsilon) \sqrt{C_3}) \sqrt{\sin(3\phi)^2 - 4\tilde k/C_3} \big) , 
\\
\begin{aligned}
R(r,\varepsilon) & = 
\sgn(\dot{r}) \tfrac{2}{45} r^{-4} \big( \sqrt{2(6E\mathstrut^\dagger -P^2)r^2 - C_3}^3 (3 (6E^\dagger -P^2)r^2 + C_3)
\\&\qquad
- \sqrt{2(6E -P^2)r^2 - C_3}^3 (3 (6E -P^2)r^2 + C_3)
\big) , 
  \end{aligned}
\end{gather}
and
\begin{equation}\label{ex2.T.Y.com}
  \bar{x}^\dagger = \bar{x} + \sgn(\dot{r}) \tfrac{1}{3} P \big(
  \sqrt{2(6E\mathstrut^\dagger -P^2)r^2 - C_3} (6E^\dagger -P^2)^{-1}
- \sqrt{2(6E -P^2)r^2 - C_3} (6E -P^2)^{-1} \big) . 
\end{equation}
\end{subequations}
This transformation group \eqref{ex2.T.group} is dynamical,
due to the presence of $P$, $C_3$, and $E$.

Last, consider $\X_{(\Psi)}$.
The dynamical transformations that it generates can be derived most easily
by a different method that uses the conserved integrals directly. 
Since $\X_{(\Psi)}E =0$, $\X_{(\Psi)}P =0$, and $\X_{(\Psi)}\Psi =0$, 
while $\X_{(\Psi)}C_{3'} =-3C_{3'}$,
this yields
\begin{equation}
  C_{3'}^\dagger = e^{-3\varepsilon}C_{3'} , 
\end{equation}
along with $E^\dagger=E$, $P^\dagger=P$, and $\Psi^\dagger=\Psi$,
where $\varepsilon\in\Rnum$ is the parameter. 
Then these transformations can be applied to the trajectory solution \eqref{ex2.trajectory}
to obtain
\begin{subequations}\label{ex2.Psi.group}
\begin{equation}\label{ex2.Psi.Y.cos}
  \sgn(\alpha)\cos(3\phi^\dagger)
  = \sgn(\dot{\phi})\sqrt{1-\tilde k/(2\tilde C_3)}\, \sin(\sgn(\dot{r})\alpha^\dagger -\Psi)
\end{equation}
where
\begin{equation}
  \cot(\alpha^\dagger) = F^\dagger(3-F^\dagger{}^2) (1-3F^\dagger{}^2)^{-1}, 
  \quad
  F^\dagger = \sqrt{(\tilde E/\tilde C_3^\dagger) r^\dagger{}^2 -1} . 
\end{equation}
In addition, $\X^\solnsp_{(\Psi)}T=0$ shows that $T^\dagger = T$, 
which combined with expression \eqref{ex2.T} leads to the transformation
\begin{equation}\label{ex2.Psi.Y.r}
  r^\dagger = \sqrt{ r^2 + 9(6E -P^2)^{-1} (\sqrt{C_3} +9\varepsilon)\varepsilon } . 
\end{equation}
\end{subequations}
Equations \eqref{ex2.Psi.Y.cos}--\eqref{ex2.Psi.Y.r}
yield the dynamical transformation generated by $\X^\solnsp_{(\Psi)}$. 

The Noether symmetry group of the system \eqref{ex2.eom}
comprises the three point transformations 
\eqref{ex2.com.transformation}, 
\eqref{ex2.time.transformation}, 
\eqref{ex2.Galilean.transformation}.
and the three dynamical transformations
\eqref{ex2.C3.group},
\eqref{ex2.T.group},
\eqref{ex2.Psi.group}.

\section{Conclusions}\label{sec:remarks}

The three examples of dynamical systems discussed in the present work 
illustrate the main results of the hybrid Lagrangian-Hamiltonian framework
developed in \Ref{Anc2026}.
For each system,
a straightforward computation of variational point symmetries in a Lagrangian setting 
leads to corresponding locally conserved integrals 
which are found to commute in the Poisson bracket
imported from the equivalent Hamiltonian setting. 
Action-angle variables are then introduced in the Lagrangian setting,
which leads to explicit integration of the Euler-Lagrange equations of motion
locally in time.

This showcases how local Liouville integrability generalizes, in a very useful way,
the more familiar concept of global Liouville integrability.
The present results fully clarify the questions about
local versus global Liouville integrability raised recently \cite{Ley}.

\section*{Acknowledgements}

S.C.A.\ is supported by an NSERC research grant.

\end{document}